\documentclass[a4paper,12pt]{article}
\pdfoutput=1
\usepackage{amsfonts,amsthm,amsmath,amssymb,graphicx,hyperref,color,youngtab}

\def\Tr{{\rm Tr\, }}
\newcommand{\Sym}{\operatorname{Sym}}

\def\Sym{\textrm{Sym}}

\newcommand{\cA}{\mathcal A}

\newcommand{\cH}{\mathcal H}

\newcommand{\cN}{\mathcal N}
\newcommand{\cO}{\mathcal O}

\newcommand{\cZ}{\mathcal Z}

\newcommand{\be}{\begin{equation}}
\newcommand{\bea}{\begin{eqnarray}}
\newcommand{\ee}{\end{equation}}
\newcommand{\eea}{\end{eqnarray}}

\newcommand{\nn}{\nonumber}

\def\mC{ \mathbb{C}}

\def\mR{ \mathbb{R}}

\begin{document}

{\LARGE{ 
\centerline{\bf  Holomorphic primary fields in free CFT4  }
\centerline{\bf  and Calabi-Yau orbifolds.   } 
}}  

\vskip.5cm 

\thispagestyle{empty} \centerline{
    {\large \bf Robert de Mello Koch
${}^{a,} $\footnote{{\tt robert@neo.phys.wits.ac.za}},
Phumudzo Rabambi,
${}^{a,} $\footnote{{\tt Phumudzo.Rabambi@students.wits.ac.za}}}
   {\large \bf Randle Rabe,
               ${}^{a,}$\footnote{{\tt 1296569@students.wits.ac.za}} }}
\centerline{{\large \bf Sanjaye Ramgoolam
               ${}^{a,b,}$\footnote{{\tt s.ramgoolam@qmul.ac.uk}}   }
                                                       }

\vspace{.4cm}
\centerline{{\it ${}^a$ National Institute for Theoretical Physics,}}
\centerline{{\it School of Physics and Mandelstam Institute for Theoretical Physics,}}
\centerline{{\it University of Witwatersrand, Wits, 2050, } }
\centerline{{\it South Africa } }

\vspace{.4cm}
\centerline{{\it ${}^b$ Centre for Research in String Theory, School of Physics and Astronomy},}
\centerline{{ \it Queen Mary University of London},} \centerline{{\it
    Mile End Road, London E1 4NS, UK}}

\vspace{1truecm}

\thispagestyle{empty}

\centerline{\bf ABSTRACT}

\vskip.2cm 
Counting formulae for general primary fields in free four dimensional conformal field theories of 
 scalars, vectors and matrices are derived.   These are specialised to count primaries which obey extremality conditions defined in terms of 
the dimensions and left or right spins (i.e. in terms of relations between the charges under the Cartan subgroup of $SO(4,2)$). 
The construction of primary fields for scalar field theory is mapped to a problem of determining multi-variable polynomials subject to 
a system  of symmetry and  differential constraints. For the extremal primaries, we give a construction in terms of 
holomorphic polynomial functions  on  permutation orbifolds, which are shown to be Calabi-Yau spaces.

\setcounter{page}{0}
\setcounter{tocdepth}{2}

\newpage

\tableofcontents

\setcounter{footnote}{0}

\linespread{1.1}
\parskip 4pt

{}~
{}~

\section{\label{sec:Introduction} Introduction  } 

In \cite{tftpap} we  showed that free scalar four dimensional conformal field theory 
can be formulated as an infinite dimensional associative algebra, admitting a decomposition into 
linear representations of  $SO(4,2)$, and equipped with a bilinear product satisfying a non-degeneracy condition. 
This algebraic structure gives a formulation of the CFT4 as a two dimensional topological field theory (TFT2)
with $SO(4,2)$ invariance, where crossing symmetry is expressed as associativity of the algebra. 
TFT2 structure had previously been identified as a unifying structure in the study of combinatorics and correlators in BPS sectors of 
$N=4$ SYM, quiver gauge theories, matrix models, tensor models, and in Feynman graph combinatorics \cite{FeynCount,QuivCalc,Belyi,TensCount}. The theme of TFT2 as a powerful unifying structure for QFT 
combinatorics was also  developed  in \cite{tftpap} in the context of counting primary fields. 
In this paper we return to a systematic study of primaries in free field theories in four dimensions. 
We consider scalar, vector and matrix models. Another motivation for the detailed construction of 
primary fields four dimensional scalar QFT is that free field calculations have been found to be useful 
in calculating the anomalous dimensions of operators at the Wilson-Fischer fixed point in the epsilon expansion \cite{Rychkov:2015naa,Basu:2015gpa,Ghosh:2015opa,Raju:2015fza,Nii:2016lpa}.

We start by developing some explicit formulae for the counting of primary fields, using characters of representations of $so(4,2)$. 
This makes extensive use of previous literature on the subject, notably \cite{Dolan05}. 
This is followed by considering the problem of constructing the primary fields.
A useful remark is that the algebraic problem of finding composite fields 
of the form 
\bea 
(\partial \cdots \partial \phi ) ( \partial \cdots \partial \phi) \cdots ( \partial \cdots \partial \phi) 
\eea
where there are $n$  $ \phi$ fields involved, can be conveniently rephrased in terms of a question about 
multi-variable polynomial functions of $ 4n$ variables : $ \Psi ( x_{\mu}^I ) $ where $ \mu $ runs over the 
space-time coordinates and $ I$ runs from $1$ to $n$. This relies on a function space realisation of the conformal algebra. 
We explain how  this function space realisation arises naturally in radial quantization. 
The question of constructing primaries, when phrased in terms of the functions $ \Psi  ( x_{\mu}^I ) $
can be viewed as a many-body quantum mechanics problem, where $F$ is a many-body wavefunction of 
$n$ particles moving on $ \mR^4$.  These many-body wavefunctions have to obey three simple conditions : 

$\bullet$ They have to obey Laplace's equation in each of the variables $ x_{\mu}^I$ for $ I = 1 \cdots n$. 

$\bullet$ They have to be invariant under the translation  $ x_{\mu}^I \rightarrow x_{ \mu}^I + a_{\mu}$, for $ \mu = 1 \cdots 4 $. 

$\bullet$ They have to be invariant under permutations $ x_{\mu}^I \rightarrow x_{\mu}^{ \sigma (I) } $ 
for any permutation $ \sigma \in S_n$. 

An infinite class of solutions  of the Laplacian condition are obtained by choosing a complex structure to 
identify $ \mR^4 = \mC^2$ so that $ x_{\mu} \rightarrow ( z , w , \bar z , \bar w )$
and considering holomorphic functions of $ z , w$. These primaries correspond to holomorphic polynomial functions on
\begin{align}\label{Orb1} 
\boxed{   ~~~~~
( \mC^2)^n /(  \mC^2 \times S_n ) ~~~~~ }
\end{align}
which can also be written as
\bea
 (\mC^n/\mC\times \mC^n/\mC)/S_n
\eea  
The modding out by $\mC^2$ is the condition of invariance under the shift while the $S_n$ invariance comes from 
the permutation symmetry. A special class of these primary fields correspond to functions of $ z $ only i.e functions on 
\bea\label{Orbsimp}  
( \mC^n) /(  \mC \times S_n ) 
\eea 
These primaries were constructed  in \cite{BHR2} using an oscillator realization of the conformal algebra, 
which is close to the differential realization used here. An extensive study of the representations of $ so(4,2)$ on function spaces 
 with emphasis on relations to quarternions, is developed in \cite{FL}. 

The association of primaries to functions on the orbifold has several interesting consequences. 
Since the holomorphic polynomial functions form a ring, and a class of primaries are in 1-1 correspondence with these 
functions, we are finding on a ring structure on this subspace of primary operators. 
This ring structure is different from the algebra structure related to the operator product expansion. 
The interplay between this product and the OPE would be an interesting subject for future study. 
The Hilbert series of the polynomial ring  (\ref{Orb1})  has a very interesting {\it palindromy property} which we prove. 
The proof relies on an interesting algebraic structure based on symmetric groups in the problem. For fixed number of primaries $n$, 
this is 
\bea 
\bigoplus_{ k , l = 0}^{ \infty} \mC( S_n ) \otimes \mC ( S_n ) \otimes \mC ( S_k ) \otimes \mC ( S_l ) 
\eea
where $ \mC ( S_n )$ is the group algebra of the symmetric group $S_n$. As recently discussed in the context of Hilbert series for moduli spaces of supersymmetric vacua of gauge theories \cite{GeomAP,SQCDAd}, 
the palindromy property of Hilbert series is indicative that the ring being enumerated is Calabi-Yau. The precise mathematical statement is due to 
Stanley \cite{stanley}. 
We show that the orbifold (\ref{Orb1})  indeed admits a unique non-singular 
nowhere-vanishing top-dimensional holomorphic form, which is inherited from the covering space. 

Our work involves an  interesting interplay between representations of $ so(4,2)$ 
and representations of symmetric groups. Let $V_+$ be the lowest weight representation corresponding to 
local operators build from derivatives acting on the field  $ \phi$. The construction of primaries
built from derivatives acting on $n$ copies of $ \phi$,  amounts to 
finding explicit formulae for the lowest weight states of  irreducible representations in the symmetrized tensor product 
$ Sym^n ( V_+)$.  If we consider the primaries which arise at dimension $n+k$,  of the class associated to the geometry (\ref{Orbsimp})
this can be mapped to a problem about multiplicities of $ S_n \times S_k $ irreps in $ V_{H}^{\otimes k  }$ where $V_H$ is the $n-1$ dimensional 
representation of $S_n$. A formula  for these multiplicities, derived in  \cite{BHR2}, is found to be useful in the study of the geometry of (\ref{Orb1}). 
The connection between representation theory of symmetric groups and that of non-compact groups has also been discussed  in 
\cite{GopGab1406} in the context if higher spin theories.

We extend this approach to primary fields to the case of vector fields in four dimensions. 
The underlying orbifold geometry for holomorphic primary fields  in this case is 
\begin{align} 
\boxed{ ~~~~~
 ( \mC^{2n}  / \mC \times \mC^{2 n}  / \mC ) / S_{n} [ S_2 ]  
~~~~~ } 
\end{align} 
The group  $ S_n[S_2]$ is the subgroup of $ S_{2n}$ which is generated by 
the $n$ pairwise permutations $ (1,2), (3,4) , \cdots , ( n-1 , n )$ along with the $n! $ permutations of these pairs. 
It is called the wreath product of $S_n$ with $S_2$. We establish the palindromy property of 
the Hilbert series in this case.  For the case of primary fields in the free theory of matrices in four dimensions, we  again find the 
  underlying orbifold geometry 
  \begin{equation} 
\boxed{ 
 ( ( \mC^{2} )^n  \times S_n ) / (\mC^2\times S_{n})=
(\mC^n/\mC\times\mC^n/\mC\times S_n)/S_n 
 }
\end{equation}
with a palindromic Hilbert series. 

The paper is organized as follows. In section \ref{sec:PolyRep} we describe a realisation of 
the conformal algebra $ so(4,2)$ in terms of differential operators acting on polynomial functions of  space-time 
coordinates $ x_{\mu}$ in $ \mR^4$ . This is related, by a duality which we explain, to the standard realization of the conformal 
algebra in terms of derivatives acting on a scalar field. In section \ref{sec:Counting} we obtain a number of useful general formulae 
for the counting of primary fields. The first step is to start from the character of the irrep $ V_+$ of $so(4,2)$ which  contains all
the local operators consisting of derivatives acting on a single scalar field. This is a function of variables $ s, x, y $ which keep track of 
dimension, left spin and right spin i.e eigenvalues of $D$ (the scaling operator) and $J_L, J_R $ (the Cartan generators for the two 
$SU(2)$'s in $ SO(4 ) = SU(2) \times SU(2)$).  We then derive a generating function for the characters 
of all the symmetrised tensor products $ Sym^n ( V_+ ) $. This is an application of the Cauchy identity. We describe a specialisation 
of these formulae relevant to what we call extremal primaries. These include the leading twist primaries studied in the context of 
deep inelastic scattering in QCD. Taylor expansion of the generating function leads to explicit results for $ n=3,4$ which take the form 
of rational functions of $ s, x, y$. In section \ref{sec:Construction} we describe the construction of the primary fields using the polynomial representations. A new counting formula for  the extremal primaries  is obtained by exploiting the 
permutation group algebras $ \mC ( S_n ) \times \mC (S_n ) \otimes \mC ( S_k ) \otimes \mC ( S_l )$ in the problem of building 
primaries from $n$ fields $\phi$ and corresponding to polynomials of degree $k$ in one holomorphic variable 
and degree $l$ in the other. This is shown to be consistent with the derivation based on the Taylor 
expansion method of the previous section. These primary fields form a ring and the counting is recognised as a Hilbert series, which encodes aspects of the 
generators and relations of the ring. This is a ring of functions of an orbifold which we identify. The counting formula based on $ S_n \times S_k \times S_l $ symmetry 
 for the extremal sector is shown to have  a palindromy property indicative of a Calabi-Yau nature of the orbifold. As further support for 
 the Calabi-Yau nature of the orbifold, we construct the explicit top-dimensional holomorphic form. In section \ref{sec:SB} we extend the results on  
counting and construction of primaries, and the underlying Calabi-Yau orbifold geometries,  to the case of a four dimensional vector model. In section \ref{sec:MM} we develop the story for the case of free four dimensional 1-matrix theory. 

This paper is the extended version of an accompanying short paper \cite{short}.

\section{\label{sec:PolyRep} Representations of  $so(4,2) $ on multi-variable polynomials }

The generators of $SO(4,2)$ form the algebra
\bea\label{so42} 
&&  [ K_{ \mu } , P_{ \nu} ] = 2 M_{ \mu \nu } - 2 D \delta_{ \mu \nu } \cr 
 && [ D , P_{ \mu} ] = P_{ \mu} \cr 
 && [ D  , K_{ \mu} ] = - K_{ \mu} \cr 
 && [ M_{ \mu \nu } , K_{ \alpha } ] = \delta_{ \nu \alpha } K_{ \mu} - \delta_{ \mu \alpha } K_{ \nu } \cr 
 && [ M_{ \mu \nu } , K_{ \alpha } ] = \delta_{ \nu \alpha } P_{ \mu} - \delta_{ \mu \alpha } P_{ \nu }
 \eea
The representations of this algebra play a central role when the constraints that conformal invariance places on the dynamics of a CFT are developed.
To develop the representation theory, one uses the fact that there is a unique primary operator ${\cal O}$ for each irrep, formed by taking products of the fundamental fields of the theory and derivatives of these fields, with each other. 
The primary operator is distinguished because its dimension can not be lowered.
Consequently, primaries are annihilated by the generator of special conformal transformations
\bea
   [K_\mu ,{\cal O} ] =0
\eea
The complete irrep is then formed by acting on the primary ${\cal O}$ with traceless symmetric polynomials in the momenta $P_\mu$.
The spectrum of dimensions of the primaries and their OPE coefficients provide a list of data that completely
determines the correlation functions of local operators.
Clearly then, it is interesting to determine the spectrum of primary operators of a conformal field theory. 
Our goal is to determine this list for the free bosonic field $\phi$ in four dimensions.
The states corresponding to $ \phi $ and its derivatives in the operator-state correspondence consists of a lowest 
weight state $ |v_+ > $
\bea 
&& D |v_+ > =  |v_+>  \cr 
&& K_{ \alpha } |v_+ > = 0 
\eea
This state obeys 
\bea 
K_{ \alpha } ( P_{\mu} P_{\mu} |0> ) = 0 
\eea
which means that $ P_{\mu} P_{\mu } |0 > $ can be set to zero to  give an irreducible representation. 
The states in this representation are of the form 
\bea 
(S^{(l)})^{\nu_1 , \nu_2 , \cdots , \nu_l }_{ \mu_1 , \mu_2 , \cdots , \mu_l } P_{\mu_1}  P_{\mu_2} \cdots P_{\mu_l } |v^+> 
\eea
where $ (S^{(l)})_{\mu_1 , \cdots , \mu_l }^{ \nu_1 , \cdots , \nu_l }$ is symmetric and traceless in both upper and lower indices.

Solving for the primaries ${\cal O}$ is a representation theory  problem of  finding the decomposition of 
the symmetrized tensor product $\Sym^n ( V_+ ) $ into irreducible representations. 
A particularly convenient realization of $ V_+ $  is in terms of harmonic polynomials.
Indeed polynomials of the form 
\bea 
(S^{(l)})^{\nu_1 , \nu_2 , \cdots , \nu_l }_{ \mu_1 , \mu_2 , \cdots , \mu_l } x_{\mu_1} \cdots x_{ \mu_l } 
\eea 
are annihilated by the Laplacian 
\bea 
{ \partial \over \partial { x_{ \alpha} }} { \partial \over \partial x_{ \alpha } } 
\eea
 hence are harmonic. 
The algebra  $so(4,2)$ is realised on these polynomials as  \cite{tftpap}
\bea\label{diffops}  
&& K_{\mu} = { \partial \over \partial x_{\mu} }\cr 
&& P_{\mu} =  ( x^2\partial_\mu -2x_\mu x\cdot\partial -2x_\mu  ) \cr 
&& D = ( x\cdot\partial +1  ) \cr 
&& M_{\mu \nu} = x_{\mu } \partial_{ \nu} - x_{\nu} \partial_{\mu} \label{pgens}
\eea
Thinking of $x_\mu$ as the co-ordinate of a particle, this is a single particle representation.
The tensor product $ V_+^{ \otimes n } $ can be realized on a many-particle space of functions 
$ \Psi ( x_{\mu}^I ) $, where $ 1 \le I \le n $ labels the particle number. 
The generators of $so(4,2)$ now include 
\begin{eqnarray}
   K_\mu = \sum_{I=1}^n {\partial\over\partial x^I_\mu}
\end{eqnarray}
\begin{eqnarray}
P_\mu =\sum_{I=1}^n \left(x^{I\rho}x^I_\rho {\partial\over \partial x^I_\mu} -2x^I_\mu x^I_\rho {\partial\over \partial x^I_\rho}
-2x^I_\mu\right)
\end{eqnarray}
along with the many-particle versions of $ D , M_{\mu \nu }$ of  (\ref{diffops}). 
In this polynomial rep,  the state of the  scalar field $\lim_{ |x| \rightarrow 0  }  \phi( x ) |0 > $ 
corresponds to the harmonic function $1$.

This polynomial representation is naturally understood  in the context of radial quantization. 
Towards this end, consider the mode expansion of the field 
\bea\label{RQradmodexp}  
\phi ( x_{ \mu } ) = \sum_{ l =0  }^{ \infty } \sum_{ m \in V_l }   a^{\dagger}_{ l ; m } Y_{ l , m } ( x )  
 + \sum_{ l = 0 }^{ \infty } \sum_{ m \in V_l } a_{ l ; m } |x|^{-2} Y_{ l ,m } ( x' ) 
\eea
The sum over $m$ is over the states of the symmetric traceless tensor irrep $V_l$ of $SO(4)$.
Acting on the vacuum, which is annihilated by the $a_{l;m}$'s but not the $a^{\dagger}_{l;m}$'s, we have the 
usual operator-state correspondence. 
For example, we find
\bea 
&& \lim_{x\rightarrow 0} \phi(x) | 0 \rangle = a^{\dagger}_{0;0} | 0 \rangle   \equiv | \phi \rangle \cr 
&& \lim_{x\rightarrow 0} \partial_{\mu}\phi(x)|0\rangle = a^{\dagger}_{1;\mu}|0\rangle
\equiv |\partial_{\mu}\phi\rangle\cr    
&&  \lim_{x\rightarrow 0}\partial_{\mu} \partial_{\mu} \phi(x) |0\rangle = 0 
\eea
The last equation above is expected because the free scalar field is a representation with null states.
It expresses the free equation of motion.
The scalar field and all its derivatives as $ x \rightarrow 0$ lead to states  in an irreducible lowest weight 
representation $V$ of $SO(4,2)$, consisting of a lowest weight state of  dimension $\Delta=1$ along with states with higher dimension. 

Let us rewrite the positive part of the  radial mode expansion 
\bea 
\phi^+(x)|0\rangle = \sum_{l=0}^{\infty}a^{\dagger}_{l;\mu_1,\cdots,\mu_l} 
( S^{(l)} )^{\mu_1,\cdots,\mu_l}_{\nu_1,\cdots,\nu_l}x^{\nu_1}\cdots x^{\nu_l}|0\rangle 
\eea
where $S^{(l)} $ is a projector, projecting to symmetric traceless tensors. 
We take $a^{\dagger}_{l;\mu_1\cdots\mu_l}$ to be symmetric and traceless in the $\mu$ indices. 
$S^{(l)} $ is symmetric and traceless in the $\mu$ as well as the $\nu$ indices.  
The operator state map identifies
\bea 
\lim_{x\rightarrow 0}\partial_{\mu_1}\cdots\partial_{\mu_l}\phi (x)|0\rangle  =
 (S^{(l)})_{\mu_1,\cdots,\mu_l}^{\nu_1,\cdots,\nu_l}a^{\dagger}_{l;\nu_1\cdots\nu_l}|0\rangle
\eea
Note that we have a duality 
\bea 
\left(\, (S^{(l)})_{\mu_1,\cdots,\mu_l}^{\nu_1,\cdots,\nu_l}a^{\dagger}_{l;\nu_1\cdots\nu_l}|0\rangle )^{\dagger} ,
\phi (x)|0\rangle\,\right) 
 &=& \langle 0|a_{l;\nu_1\cdots\nu_l} (S^{(l)})_{\mu_1,\dots,\mu_l}^{\nu_1,\cdots,\nu_l}\phi (x) |0\rangle \cr 
 &=& ( S^{(l)})_{\mu_1,\cdots,\mu_l}^{\alpha_1,\cdots,\alpha_l} x_{\alpha_1} \cdots x_{\alpha_l} 
\eea
where we have used the projector property of $S^{(l)}$. 
Unpacking this a little, if we apply $ \partial_{\mu}$ to the local operator, go to zero to get the corresponding 
state and then do the duality, we will get a new polynomial as the outcome
\bea 
  \lim_{x\rightarrow 0}\partial_{\mu} \partial_{\mu_1}\cdots\partial_{\mu_l}\phi ( x ) |0\rangle =
 (S^{(l)})_{\mu,\mu_1,\cdots ,\mu_l}^{\nu,\nu_1,\cdots,\nu_l}a^{\dagger}_{l+1;\nu,\nu_1\cdots\nu_l}|0\rangle  
\eea
If we take the overlap of this with $\phi (x)|0\rangle $ then we get 
\bea 
 (S^{(l)})_{\mu, \mu_1, \cdots , \mu_l  }^{ \nu, \nu_1, \cdots , \nu_l }  x_{ \nu } x_{ \nu_1} \cdots x_{ \nu_l }
\eea
This polynomial of degree one higher is related to the previous polynomial by applying 
$P_{\mu}=\left(x^2\partial_{\mu}-2 x_{\mu}(x\cdot\partial +1)\right)$.
We have the following identifications between operators and states, and then states and polynomials
\bea 
&& \cO  \rightarrow | \cO \rangle \rightarrow P_{ \cO } ( x ) \cr 
&& \partial_{\mu} \cO \rightarrow | \partial_{\mu} \cO \rangle \rightarrow P_{\mu} P_{ \cO } ( x )  
\eea
This provides a concrete correspondence between applying $\partial_{\mu}$ to local operators made from 
a scalar, and applying $P_{\mu}$ as the dual differential operator on dual polynomials. 

Primaries in the free theory are given by acting with traceless symmetric polynomials in momenta, on the scalar field.
Tracelessness is often implemented\cite{Dobrev:1975ru,Costa:2011mg} by using variables
 $z\cdot x^I = z^\mu x^I_\mu$ with $z^\mu$ a null vector, i.e. $z^\mu z_\mu =0$.
Thanks to the fact that $z^\mu$ is null, any polynomial in $z\cdot x^I$ automatically gives a traceless 
symmetric polynomial in $x^I_\mu$ after the $z^\mu$s are stripped away.
In what follows we will solve the algebraic primary problem, to obtain a polynomial that corresponds to the primary.
To obtain the primary operator written in terms of the original scalar field, we need to translate between the polynomials
and operators. 
For the current polynomials, the translation between polynomials and operators is
\begin{eqnarray}
  (z\cdot\partial )^k\phi \leftrightarrow (-1)^k 2^k k! (z\cdot x)^k\label{polytranslate}
\end{eqnarray}
This construction is convenient because of its simplicity.
However, it is not completely general, since there are primary operators that are not symmetric in
their indices and hence can't be represented as a polynomial in $z\cdot x$.
The general discussion makes use of projectors that project from symmetric tensors to traceless symmetric tensors. 
It is useful to consider a concrete example.
The tensors of ranks 2 and 3 are given by
\begin{eqnarray} 
(S^{(2)})_{\mu\nu}^{ \alpha \beta} & = &  \delta^{\alpha }_{\mu} \delta^{\beta}_{\nu} - {1\over 4}\delta_{\mu\nu} \delta^{\alpha\beta}\cr 
(S^{(3)})_{\mu\nu\rho}^{\alpha\beta\gamma}& = &\delta^{\alpha}_{\mu}\delta^{\beta}_{\nu}\delta^{\gamma}_{\rho} 
   - {1\over 6} (\delta_{\mu\nu} \delta^{\alpha\beta}\delta_\rho^{\gamma}
+\delta_{\mu\rho}\delta^{\alpha\gamma}\delta_\nu^\beta
+ \delta_\mu^\alpha \delta^{ \beta \gamma}\delta_{\nu\rho}) 
\cr
&&
\end{eqnarray}
These operators are projectors in the Brauer algebra of tensor operators that commute with $SO(4)$\cite{brauer}
\begin{eqnarray} 
S^{(2)} & = &  1 - {  C_{ 12} \over 4 } \cr  
S^{(3)} & = & 1 - { 1 \over 6  } \left ( C_{ 12} + C_{ 13} + C_{23} \right ) 
\end{eqnarray}
The terms correcting the $1$ above subtract off the trace of the tensors they act on.
They satisfy 
\begin{eqnarray}
\label{prod}  
(S^{(n)} )^2 P_{n}=S^{(n)} P_{n} \label{projprop}
\end{eqnarray}
where $P_n$ projects onto the totally symmetric polynomials of degree $n$
\begin{eqnarray} 
P_{ n} = { 1 \over n! } \sum_{ \sigma \in S_n } \sigma 
\end{eqnarray}
The multiplication (\ref{prod})  is in the Brauer algebra, where loops are assigned the value of $4$.
These elements of the Brauer algebra are completely determined by the projector property (\ref{projprop}) and the  property that they start with $1$. 
In general
\begin{eqnarray} 
P_{\mu_1}\cdots P_{\mu_k }\cdot  1= (-1)^k  {2^k   k!} (S^{(k)} )_{\mu_1\cdots\mu_k }^{\nu_1\cdots\nu_k }x_{\nu_1} \cdots x_{\nu_k } 
\end{eqnarray}
The above factor is easily obtained by deriving a recursion formula.
Note that the term $x^2\partial_{\mu}$ does not raise the rank of the tensor. 
The other two terms both raise the rank by one, which then leads to the recursion relation. 
In the many-particle realization such a traceless polynomial made of the $I$'th coordinates corresponds to 
derivatives acting on the $I$'th copy of  $\phi$ in a sequence of $n $ of these. 

To construct primaries using $n$ scalar fields we consider a multi-particle system with $x^I_\mu$ the coordinates
of the $n$ particles.
Primaries at dimension $n+k$ are obtained by allowing $k$ derivatives to act on the $n$ fields.
In the dual polynomial language, states at dimension $n+k$ in $V^{\otimes n}$ correspond to polynomials 
in $x^{I}_{\mu} $ of degree $k$. 
Primaries at dimension $n+k$ correspond to degree $k$ polynomials $\Psi(x^I_{\mu})$  that obey the conditions
\begin{eqnarray}\label{constraints}  
&&K_{\mu}\Psi = \sum_{I}\frac{\partial}{\partial x^I_{\mu}}  \Psi =0 \cr 
&&{\cal L}_{I}  \Psi = \sum_{\mu} \frac{\partial}{\partial x^I_{\mu}}\frac{\partial}{\partial x^I_{\mu}}\Psi  =0\cr
&&   \Psi ( x_{\mu}^{ I } ) = \Psi ( x_{ \mu}^{ \sigma (I) } ) 
\end{eqnarray}
The first condition above is the familiar condition that the special conformal generator annihilates primary
operators.
The second condition implements the free scalar equation of motion which implies that the image of states like
$P_{\mu}^{I} P_{\mu}^{I}$, with only $\mu$ summed, in the Fock space, is zero. 
This null state appears because the dimension of free scalar field saturates a unitarity bound.
To see that the second constraint is indeed implementing the equation of motion, note that with the second of
(\ref{pgens}) we can calculate
\bea 
P_{\mu} P_{\mu} = x^4 \partial_{\mu} \partial_{\mu} 
\eea
Simplifying the product of differential operators, it is simple to verify that terms like 
$x^2$, $ x^2 x\cdot\partial $ and $ x^2 x_{\mu} x_{\nu} \partial_{\mu} \partial_{\nu} $ cancel out.  
The final condition in (\ref{constraints}) above ensures that our polynomials are $S_n$ invariant.
By constructing $S_n$ invariant polynomials, we are implementing the bosonic statistics of the scalar field.

In what follows we will focus on primaries (and hence polynomials) that transform in a definite representation of 
the $SO(4)=SU(2)\times SU(2)$ subgroup of $SO(4,2)$.
To make the $SO(4)$ transformation properties of the polynomials more transparent, 
our construction makes use of the complex coordinates
\begin{eqnarray}
  z=x_1+ix_2\qquad w=x_3+ix_4\cr
 \bar z=x_1-ix_2\qquad \bar w=x_3-ix_4
\end{eqnarray}
This amounts to choosing an isomorphism between $\mathbb{R}^4$ and $\mC^2 = \mathbb{C}\times\mathbb{C}$. 
In our conventions, these coordinates have the following $(j^3_L,j^3_R)$ charge assignments
\begin{eqnarray}\label{complex-coords} 
&&z\leftrightarrow ({1\over 2},{1\over 2})\qquad
\bar z\leftrightarrow (-{1\over 2},-{1\over 2})\cr
&&w\leftrightarrow ({1\over 2},-{1\over 2})\qquad
\bar w\leftrightarrow (-{1\over 2},{1\over 2})
\end{eqnarray}
We will construct a class of   primaries corresponding to holomorphic polynomial functions on the orbifold 
\begin{eqnarray} 
( \mC^2 )^n  / ( \mC^2 \times S_n)  
\end{eqnarray} 
The division by $\mC^2 $  is a consequence of the first of (\ref{constraints}).
These will not form the complete set of primaries but a well-defined subspace of primaries, which we will call {\it extremal}. 
Before explaining this construction in more detail we show, in the next section, how characters of $so(4,2)$ 
representations can be used to get a complete counting of general primaries built from $n$ fields. We will then specialize 
to the extremal primaries.

\section{\label{sec:Counting} Counting with $so(4,2)$ characters }

In this section our goal is to enumerate the $SO(4,2)$ irreducible representations appearing 
among the composite fields made out of $n=2,3, \cdots$ fundamental fields. 
These multiplicities will, for example, compute the spectrum of primary operators in the free CFT$_4$.
This enumeration entails decomposing, into irreducible representations, the symmetrized tensor product $\Sym^n (V_+)$,
where $V_+=D_{[1,0,0]}$ in the notation of \cite{Dolan05}. 
The three integer labels in $D_{[\Delta,j_L,j_R]}$ are the dimension and two $SO(4)$ spins.
After obtaining a general formula in terms of an infinite product, we specialize to primaries that obey extremality
conditions, that relate their dimension to their spin. For these primaries using results from \cite{BHR2}, we
find simple explicit formulas for the counting. 

\subsection{General Counting Formula}

Consider a matrix $M$ belonging to any matrix representation $R$ of $SO(4,2)$.
A key result for the analysis of this section is
\bea
{1\over {\rm det}(1-tM)}=\sum_{n=0}^\infty t^n\chi_{\Sym^n (R)}(M)
\label{identity}
\eea
This is a special case of the Cauchy identity which states that 
\bea 
\prod_{ i=1}^N \prod_{ j =1 }^M 
{ 1 \over ( 1 - t x_i y_j ) } = \sum_{ n=0}^{ \infty} \sum_{ R \vdash n } \chi_R ( x ) \chi_R ( y )  
\eea
where $\chi_R $ is a Schur polynomial in the $ N $ variables $x_i $ and the $M$ variables $ y_j $, 
labelled by a Young diagram $R$ with $n$ boxes and height no larger that the minimum of $ M , N$.
When one of these variables is $1$, then we sum over single-row Young diagrams.  
This formula (\ref{identity}) is  also easily proved by using the identity (this is just a statement of Wick's theorem)
\bea
(I_n)^{j_1\cdots j_n}_{i_1\cdots i_n}
&=&{1\over \pi^N}\int\prod_{i=1}^N dx_i dy_i e^{-\sum_k z_k\bar z^k}
{1\over n!} z_{i_1}\cdots z_{i_n} \bar z^{j_1}\cdots \bar z^{j_n}\cr
&=& {1\over n!}\sum_{\sigma\in S_n}\delta^{j_1}_{i_{\sigma (1)}} \delta^{j_2}_{i_{\sigma (2)}}\cdots
\delta^{j_n}_{i_{\sigma (n)}}
\eea
to evaluate
\bea
{1\over \pi^N}
\int\prod_{i=1}^N dx_i dy_i e^{-\sum_{i,j} z_i(\delta^i_j -tM^i_j)\bar z^j}={1\over {\rm det}(1-tM)}\label{frst}
\eea
Now, apply (\ref{identity}) to the case that
\bea
  M=s^D x^{J_{3,L}} y^{J_{3,R}}
\eea
and specialize to the representation $V_+$ spanned by the free scalar and all the derivatives acting on it. 
Here we have chosen $D,J_{3,L},J_{3,R}$ to span the Cartan subalgebra of $SO(4,2)$.
It is straight forward to see that
\bea
{1\over {\rm det}(1-tM)}=\prod_{q=0}^\infty\prod_{a=-{q\over 2}}^{q\over 2}\prod_{b=-{q\over 2}}^{q\over 2}
{1\over 1-t s^{q+1}x^a y^b}
\eea
This generating function of the characters of the symmetrized tensor products of 
the free scalar representation will be denoted by  ${\cal Z}(t, s,x,y) $. So we have 
\begin{eqnarray}
{\cal Z}(t, s,x,y)=
\prod_{q=0}^\infty\prod_{a=-{q\over 2}}^{q\over 2}\prod_{b=-{q\over 2}}^{q\over 2}
{1\over 1-t s^{q+1}x^a y^b}=\sum_{n=0}^\infty t^n \chi_{\Sym^n (V_+)} (s,x,y)
\label{forchi}
\end{eqnarray}
where we have denoted $\chi_{\Sym^n  (V_+)} (M)$ by $\chi_{\Sym^n  (V_+)} (s,x,y)$.
The characters for $\Sym^n  (V_+)$ follow by developing the infinite product above
in a Taylor series in $t$.
The decomposition of $\Sym^n  (V_+ )$ into irreps is now achieved by writing 
$\chi_{\Sym^n  (V_+)} (s,x,y)$ as a sum of characters $\chi_{[\Delta,j_1,j_2]}(s,x,y)$ of $M$, in
the the irrep of dimension $\Delta$ and spins $j_1,j_2$ 
\bea
   \chi_{\Sym^n (V_+)}(s,x,y)=\sum_{\Delta,j_1,j_2}N_{[\Delta,j_1,j_2]}\chi_{[\Delta,j_1,j_2]}(s,x,y)
\eea
The coefficients $N_{[\Delta,j_1,j_2]}$ are non-negative integers, counting the number of times irrep
$\cA_{ [\Delta,j_1,j_2]} $  (in the notation of \cite{Dolan05}) appears in $Sym^n (V_+)$). 
If we restrict to the case that $n\ge 3$, the only characters $\chi_{[\Delta,j_1,j_2]}(s,x,y)$ which contribute 
are labeled by dimensions $\Delta$ that do not saturate the unitarity bound and hence do not have any null states.
In this case we have\cite{Dolan05}
\bea
   \chi_{[\Delta,j_1,j_2]}(s,x,y)=
{s^\Delta\chi_{j_1}(x)\chi_{j_2}(y)\over (1-s\sqrt{xy})(1-s\sqrt{x\over y})(1-s\sqrt{y\over x})(1-{s\over\sqrt{xy}})}
\eea
It is useful to define 
\bea\label{eq:partition}
Z_n (s,x,y)&\equiv& \sum_{\Delta,j_1,j_2}N_{[\Delta,j_1,j_2]}s^\Delta\chi_{j_1}(x)\chi_{j_2}(y)
\eea
It follows that 
\bea 
Z_n ( s, x,  y ) =
(1-s\sqrt{xy})(1-s\sqrt{x\over y})(1-s\sqrt{y\over x})(1-{s\over\sqrt{xy}}) ~ \chi_{\Sym^n(V)}(s,x,y) 
\eea
The right hand side of this last equation is precisely a sum of (products of) $SU(2)$ characters, so we can treat 
this, following \cite{spradlin}, using the orthogonality of $SU(2)$ characters.
The result is most easily stated in terms of the generating function
\bea\label{eq:gen}
   G_n(s,x,y)=\sum_{d=0}^\infty\sum_{j_1,j_2}N_{[n+d,j_1,j_2]}s^{n+d} x^{j_1} y^{j_2}
\eea
which is given by
\bea
  G_n(s,x,y)=\left[ (1-{1\over x})(1-{1\over y})Z_n (s,x,y)\right]_{\ge}\label{forG}
\eea
where the subscript $\ge$ is a notation to indicate that the above function should first be expanded
as a Laurent series in both $x$ and $y$, and then negative powers of $x$ and $y$ should be discarded.
The infinite product in the above formula makes it difficult to evaluate $G_n(s,x,y)$ in closed form.
For that reason, in the next section, we focus on specific classes of primaries for which $G_n(s,x,y)$ can be evaluated.

To end this section let us explain how the above derivation is generalized when irreps that include null states appear
in the tensor product $\Sym^n (V_+)$.
This is the case when $n=2$.
Naively computing $G_{2}(s,x,y)$ using (\ref{forG}), we obtain the following terms
\bea
G_{2}(s,x,y) &=& s^2+s^4 x y-s^5 \sqrt{x} \sqrt{y}+s^6 x^2 y^2-s^7 x^{3/2} y^{3/2}+\cdots
\eea
The negative coefficients in the above expansion show this answer is manifestly wrong.
The problem is that we have some null states that have not been removed correctly.
There are two types of primaries that appear in the above sum.
We have a primary with $\Delta =2$ and $j_1=j_2=0$ and primaries with $\Delta = 2+2j$ and $j_1=j_2=j$
for $j=1,2,3,...$.
The condition for a short multiplet\cite{Minwalla:1997ka} is that $\Delta = f(j_1)+f(j_2)$ 
with $f(j)=0$ if $j=0$ or $f(j)=j+1$ if $j>0$.
The primary with $\Delta =2$ and $j_1=j_2=0$ is not short and nothing needs to be subtracted.
The primaries with $\Delta = 2+2j$ and $j_1=j_2=j$ are short irreps and hence have null states.
These null states (and their descendants) must be removed.
To understand how this is done, note that the primary with $\Delta = 2+2j$ and $j_1=j_2=j$ is a conserved
higher spin current $J^{\mu_1\mu_2\cdots\mu_j}$ and the null state is nothing but the conservation law
\bea
   \partial_\mu  J^{\mu\mu_2\cdots\mu_j}=0
\eea
The null state thus has $\Delta = 3+2j$ and $j_1=j_2=j-{1\over 2}$ and so the subtraction of null states is achieved
by removing the primary that does not need to be subtracted, dividing by $1-s/\sqrt{xy}$  and then putting the original
primary back in.
In the end we have
\bea
  G_2 (s,x,y)&=&\left[ 
(1-{1\over x})(1-{1\over y})\left( Z_2 (s,x,y) -s^2\right)
{1\over 1-{s\over\sqrt{xy}}}
\right]_{\ge}+s^2\cr
&=&\sum_{j=0}^\infty  s^{2+2j}x^j y^j
\eea
This is indeed the correct result.

\subsection{Counting the Leading Twist Primaries}

Consider the leading twist primaries, which have quantum numbers $[\Delta,j_1,j_2]=[n+q,{q\over 2},{q\over 2}]$.
Each such primary operator comes in a complete spin multiplet of $(q+1)^2$ operators.
Choosing the operator with highest spin corresponds to studying polynomials constructed using only the single
complex variable $z$, as we can see from (\ref{complex-coords}). 
This corresponds to the fact that all primaries are constructed using a single component  $P_z$ of the momentum
four vector operator.
We will now count the leading twist primaries by counting this highest spin operator in each multiplet.
Denote the corresponding generating function by $G_n^{\rm max} (s,x,y)$.
To determine this generating function we will modify the above results in three ways:

\begin{itemize}

\item[1.] We modify the formula (\ref{forchi}) by replacing $\chi_{\Sym^n (V)}(s,x,y)$ with a new function
$\chi^{\rm max}_n (s,x,y)$, and we keep only the highest spin state in the product
\bea
\prod_{q=0}^\infty
{1\over 1-t s^{q+1}x^{q\over 2} y^{q\over 2}}=\sum_{n=0}^\infty t^n \chi^{\rm max}_n (s,x,y)
\eea

\item[2.] 
The leading twist primaries are all constructed using a single component of the momentum, that raises both the 
left and right spin maximally. Consequently in (\ref{forG}) we keep only the factor that corresponds to this component 
of the momentum, which amounts to replacing
\bea
(1-s\sqrt{xy})(1-s\sqrt{x\over y})(1-s\sqrt{y\over x})(1-{s\over\sqrt{xy}})\to (1-s\sqrt{xy})
\eea

\item[3.] For each spin multiplet we keep only 1 state so there is no longer any need to replace the multiplet of spin
states by a single state when we count. 
Thus in (\ref{forG}) we replace
\bea
(1-{1\over x})(1-{1\over y})\to 1
\eea

\end{itemize}

The final result is
\bea
  G_n^{\rm max} (s,x,y)= \chi^{\rm max}_n (s,x,y) (1-s\sqrt{xy})
\eea
%
In this formula we don't need to track the dependence on $x$ and $y$ since for this class of primaries, once
$n$ and the dimension of the operator is specified, the spins are determined.
For simplicity then, we will study
\bea
\sum_{n=0}^\infty t^n G^{\rm max}_n (s)
=\sum_{n=0}^\infty t^n (1-s) \chi^{\rm max}_n (s)
=(1-s)\prod_{q=0}^\infty
{1\over 1-t s^{q+1}}
\eea
To extract $G^{\rm max}_n (s)$, we need to develop the infinite product above in a Taylor series in $t$.
To do this we introduce the functions
\bea
F(t,s)=\prod_{q=0}^\infty {1\over 1-t s^{q+1}}\qquad
{\partial F\over\partial t}=f_1 F
\qquad f_k={\partial^{k-1}f_1\over \partial t^{k-1}}
\eea
It is straightforward to find $F(0,s)=1$ and
\bea
f_k(t,s)=(k-1)!\,\, \sum_{a=0}^\infty {s^{ka+k}\over (1-t s^{a+1})^k}\qquad
   f_k(0,s)=(k-1)!{s^k\over 1-s^k}
\eea
Using these quantities, we have
\bea
   {\partial^n F\over\partial t^n}=
\sum_{n_1,\cdots, n_q}\sum_{k_1,\cdots, k_q}
{(n_1k_1+\cdots+n_q k_q)!\over n_1!\cdots n_q!(k_1!)^{n_1}\cdots (k_q!)^{n_q}}
f_{k_1}^{n_1}\cdots f^{n_q}_{k_q}\delta_{n,n_1k_1+\cdots n_q k_q}\,\, F
\eea
Inserting the formulas for the $f_k$'s we have
\bea
   {\partial^n F\over\partial t^n}\Big|_{t=0}&=&
\sum_{n_1,\cdots, n_q}\sum_{k_1,\cdots, k_q}
{(n_1k_1+\cdots+n_q k_q)!\over n_1!\cdots n_q!\, k_1^{n_1}\cdots k_q^{n_q}}
\big({s^{k_1} \over 1-s^{k_1}}\big)^{n_1}\cdots 
\big({s^{k_q}\over 1-s^{k_q}}\big)^{n_q}
\delta_{n,n_1k_1+\cdots n_q k_q}\cr
&=&\sum_{n_1,\cdots, n_q}\sum_{k_1,\cdots, k_q}
{n! s^n\over n_1!\cdots n_q!\, k_1^{n_1}\cdots k_q^{n_q}}
\big({1 \over 1-s^{k_1}}\big)^{n_1}\cdots 
\big({1\over 1-s^{k_q}}\big)^{n_q}
\delta_{n,n_1k_1+\cdots n_q k_q}\cr
&&\label{genformI}
\eea
Notice that this is a sum over conjugacy classes of $S_n$.
The conjugacy class collects permutations with $n_q$ $k_q$-cycles.
This interpretation follows because the coefficient
\bea
{n!\over n_1!\cdots n_q!\, k_1^{n_1}\cdots k_q^{n_q}}
\eea
is the order of the conjugacy class. 
There is a factor of $(1-s^k)^{-1}$ for each $k$-cycle in the permutation.
Here are a few motivational examples
\bea
   {\partial F\over\partial t}\Big|_{t=0}={s\over 1-s}\qquad
   {\partial^2 F\over\partial t^2}\Big|_{t=0}=
{s^2\over (1-s)^2}+{s^2\over 1-s^2}={2s^2\over (1-s)(1-s^2)}
\eea
\bea
   {\partial^3 F\over\partial t^3}\Big|_{t=0}=
{s^3\over (1-s)^3}+3{s^2\over (1-s)(1-s^2)}+{2s^3\over 1-s^3}={6s^3\over (1-s)(1-s^2)(1-s^3)}
\eea
It is easy to identify the above expressions: Recall the lowest weight discrete series irrep of $SL(2)$, denoted
$V_1$, has character
\bea
   \chi_1(s)={\rm Tr}_{V_1}(s^{L_0})={s\over 1-s}
\eea
It then follows that ($P_{[n]}$ projects onto the symmetric irrep i.e. a single row of $n$ boxes)
\bea
   {\partial F\over\partial t}\Big|_{t=0}&=&{s\over 1-s}\cr\cr
                                                           &=&\chi_1 (s)
\eea
\bea
  {1\over 2!} {\partial^2 F\over\partial t^2}\Big|_{t=0}&=&
{s^2\over 2(1-s)^2}+{s^2\over 2(1-s^2)}={\rm Tr}(P_{[2]} s^{L_0})\cr\cr
&=&{\rm Tr}_{Sym(V_1^{\otimes 2})}(s^{L_0})
\eea
\bea
 {1\over 3!}{\partial^3 F\over\partial t^3}\Big|_{t=0}&=&
{s^3\over 3! (1-s)^3}+{3s^2\over 3! (1-s)(1-s^2)}+{2s^3\over 3!(1-s^3)}={\rm Tr}(P_{[3]} s^{L_0})\cr\cr
&=&{\rm Tr}_{Sym(V_1^{\otimes 3})}(s^{L_0})
\eea
This interpretation follows for general $n$ as proved in (\ref{genformI}).
Thus the general formula is
\bea
  {1\over n!} {\partial^n F\over\partial t^n}\Big|_{t=0}={\rm Tr}(P_{[n]} s^{L_0})=
{s^n\over (1-s)(1-s^2)(1-s^3)\cdots (1-s^n)}
\eea
where the last equality follows from  eqn (47) of BHR2, where these $SL(2)$ sector primaries were studied 
in the language of oscillators. 
Consequently we have
\bea
G^{\rm max}_n (s)={(1-s)\over n!}   {\partial^n F\over\partial t^n}\Big|_{t=0}=
{s^n\over (1-s^2)(1-s^3)\cdots (1-s^n)}\label{Polyinz}
\eea
Note the close connection between counting leading twist primaries and the multiplicities of 
$V^{SL(2)}_{\Lambda =n+k}\otimes V^{S_n}_{[n]}$, which is given by the coefficient of $q^k$ in
\bea
\prod_{i=2}^n {1\over 1-q^i}
\eea
The result (\ref{Polyinz}) was also recently obtained in \cite{Roumpedakis:2016qcg}.

There are three other sectors of primaries that are closely related to this one: polynomials in $\bar z$ correspond
to primaries of the form $[n+q,-q,-q]$, polynomials in $w$ to primaries of the form $[n+q,q,-q]$ and
polynomials in $\bar w$ to primaries of the form $[n+q,-q,q]$.

\subsection{\label{sec:LS} Extremal Primaries}

We now come to a more general class of primaries with charges 
\bea 
 \Delta = n + q ~~ ; ~~   J_3^L  = { q \over 2 }   
\eea
The charge $ J_3^R $, which is part of $SU(2)_R$, is not constrained. These primary operators belong to complete multiplets of $SU(2)_R$.  They correspond  polynomials constructed using 
the pair of complex variables $z_I ,w_I$. This is clear from inspection of the charges in (\ref{complex-coords}).
Translating from the polynomial representation back to the usual scalar field representation, this corresponds to the 
fact that all primaries are constructed using only two components of the momentum four vector operator.
The two components are complex linear combinations of the (hermitian) $P_\mu$.
Arguing as we did in the previous section, we introduce a generating function $G^{z,w}_n(s,x,y)$, which
is now given by
\begin{eqnarray}
G^{z,w}_n(s,x,y)=\left[\left(1-{1\over y}\right)Z^{z,w}_n(s,x,y)\right]_{\ge} \label{zwcount}
\end{eqnarray}
where $Z_n(s,x,y)$ is obtained from
\begin{eqnarray} 
\prod_{q=0}^{\infty}\prod_{m=0}^{q}{1\over (1- t s^{q+1}x^{q/2}y^{m-q/2})}=\sum_{n=0}^{\infty}t^n 
\chi_{n} (s,x,y) 
\label{zwprod}
\end{eqnarray}
\bea
Z^{z,w}_n(s,x,y)=(1-s\sqrt{xy})(1-s\sqrt{x/y})\chi_n(s,x,y)\label{forzzw}
\eea
The two brackets multiplying $Z_n(s,x,y)$ in (\ref{zwcount}) is a consequence of the fact that two components of 
the momentum four vector are used when constructing the primaries.
From (\ref{zwprod}) it is clear that we are selecting the state from the $J_{3,L}$ multiplet 
(recorded using the variable $x$) with the highest spin. 
The product over $m$ in (\ref{zwprod}) indicates that all the states in the $J_{3,R}$ multiplet are counted.
The factor of $(1-1/y)$ as well as the instruction (indicated with the subscript $\ge$ in (\ref{zwcount})) to keep 
only positive powers of $y$ ensures that we count each $SU(2)_R$ spin multiplet once.
It is clear that the expansion of (\ref{zwcount}) has only positive powers of $x$.
This is a consequence of the fact that we kept only one state from each $SU(2)_L$ multiplet.

It is again possible to derive closed expressions for the generating functions $Z^{z,w}_n(s,x,y)$ 
and $G^{z,w}_n(s,x,y)$. Introduce the functions
\bea
F_2 (t,s,x,y)
&=&\prod_{q=0}^\infty\prod_{m=0}^q{1\over 1-\, t\, s^{q+1}x^{q\over 2}y^{m-{q\over 2}}}
=\sum_{n=0}^\infty t^n \chi_n (s,x,y)\cr
{\partial\over\partial t}F_2 (t,s,x,y)
&=&\sum_{q=0}^\infty\sum_{m=0}^q 
{s^{q+1}x^{q\over 2}y^{m-{q\over 2}}\over 1-t s^{q+1}x^{q\over 2}y^{m-{q\over 2}}}F_2(t,s,x,y)
\equiv f_1(t,s,x,y) F_2(t,s,x,y)\cr
f_k(t,s,x,y)&\equiv&{\partial^{k-1} f_1\over\partial t^{k-1}}
=(k-1)!\sum_{q=0}^\infty\sum_{m=0}^q 
{s^{kq+k}x^{qk\over 2}y^{km-{kq\over 2}}\over (1-t s^{q+1}x^{q\over 2}y^{m-{q\over 2}})^k}
\eea
It is simple to establish that $F_2 (0,s,x,y)=1$ and
\bea
f_k(0,s,x,y) = s^k (k-1)!{1\over 1-s^k x^{k\over 2}y^{k\over 2}}{1\over 1-s^k x^{k\over 2} y^{-{k\over 2}}}
\eea
Exactly as above we have
\bea
   {\partial^n F_2\over\partial t^n}\Big|_{t=0}=
\sum_{n_1,\cdots, n_q}\sum_{k_1,\cdots, k_q}
{(n_1k_1+\cdots+n_q k_q)!\over n_1!\cdots n_q!(k_1!)^{n_1}\cdots (k_q!)^{n_q}}
f_{k_1}^{n_1}\cdots f^{n_q}_{k_q}\delta_{n,n_1k_1+\cdots n_q k_q}
\eea
Inserting the formulas for the $f_k$'s and streamlining the notation by using $a=s\sqrt{xy}$ and $b=s\sqrt{x\over y}$,
we find
\bea\label{TaylExpForm}
{1\over n!}{\partial^n F_2\over\partial t^n}\Big|_{t=0}&=&
\sum_{n_1,\cdots, n_q}\sum_{k_1,\cdots, k_q}
{s^n\over n_1!\cdots n_q!\, k_1^{n_1}\cdots k_q^{n_q}}
\big({1 \over (1-a^{k_1})(1-b^{k_1})}\big)^{n_1}\cr
&&\cdots \big({1\over (1-a^{k_q})(1-b^{k_q})}\big)^{n_q}
\delta_{n,n_1k_1+\cdots n_q k_q}\cr
&=&\chi_n(s,x,y)
\eea
The expression for $Z_n(s,x,y)$ now follows from (\ref{forzzw}).

It is not easy to proceed for general $n$, but it is straight forwards to obtain explicit formulas once a
specific $n$ is chosen.
For example, the final result for $n=3$ fields is
\bea
Z^{z,w}_3 (s,x,y)=
{s^3\left(s^6x^3+s^4x^2+s^2x+1+s^3x^{3\over 2}\left(\sqrt{y}+{1\over\sqrt{y}}\right)\right)\over
(1-s^2 xy)(1-s^3(xy)^{3\over 2})(1-s^2{x\over y})(1-{s^3x^{3\over 2}\over y^{3\over 2}})}
\eea
To extract spin multiplets, we need to compute
\bea
G^{z,w}_3(z,w)=
\left[ Z_3 (s,x,y) \left(1-{1\over y}\right)\right]_{\ge}
={1\over 2\pi i}\oint_C dz
{\left(1-{1\over z^2}\right)Z_3(s,x,z^2)\over z-\sqrt{y}}
\eea
The contour $C$ must have a radius larger than $\sqrt{y}$. 
We assume that $s$, $x$ and $y$ are all less than one so that the expansion of $Z^{z,w}_3 (s,x,y)$ converges.
Thus, we can take $C$ to be the unit circle.
The integrand has poles at $z=\pm s \sqrt{x}$, $z=\sqrt{y}$, $z=\pm\frac{1}{s \sqrt{x}}$,  
$z=-\frac{s \sqrt{x}}{2}\left(1\pm i\sqrt{3}\right)$ and $z=-\frac{ \left(1\pm i\sqrt{3}\right)}{2 s \sqrt{x}}$.
To compute the integral we need to pick up the residues from the poles at $z=\pm s\sqrt{x}$, 
$z=\sqrt{y}$, and $z=-\frac{s \sqrt{x}}{2}\left(1\pm i\sqrt{3}\right)$. 
We obtain
\bea
G^{z,w}_3(z,w)=
{s^3 (1-s^{10} x^{5}y^{3})
\over (1-s^4 x^2)(1-s^3\sqrt{x^3y^3})(1-s^2 xy)(1-s^5 x^{5\over 2}y^{3\over 2})}
\label{nis3zw}
\eea
It is easy to check, using mathematica, that this expression has the correct expansion.
The check tests that the expansion, as a polynomial about $s=0$, of the above generating function 
matches the counting following from the expansion of the function appearing in (\ref{forG}).

Consider next the final result for $n=4$ fields, which is
\bea
Z_4^{z,w}(s,x,y)&=&{1\over 4!}{\partial^4 F_2\over\partial t^4}\Big|_{t=0}\cr
&&=\frac{s^4 Q(s,x,y)}{\left(s^2 x-y\right)^2 \left(1-s^2 x y\right)^2 \left(s^2 x+y\right) 
\left(-s^3 x^{3\over 2}+y^{3\over 2}\right) \left(1+s^2 x y\right) \left(1-s^3 x^{3\over 2} y^{3\over 2}\right)}\cr\cr
&&\cr
&&\eea
\bea
Q(s,x,y)=y^{7\over 2} \big(y+s^2 x y+s^{10} x^5 y+s^{12} x^6 y+s^3 x^{3\over 2} y^{1\over 2}
\left(1+y\right)+s^5 x^{5\over 2} y^{1\over 2} \left(1+y\right)\cr
+s^7 x^{7\over 2} y^{1\over 2} \left(1+y\right)+s^9 x^{9\over 2} y^{1\over 2} 
\left(1+y\right)+s^4 x^2 \left(1+y\right)^2+s^6 x^3 \left(1+y\right)^2+s^8 x^4 \left(1+y\right)^2\big)\cr
\eea
To extract spin multiplets, we again need to compute
\bea
G_4^{z,w}(s,x,y)=\left[ Z_4^{z,w}(s,x,y) \left(1-{1\over y}\right)\right]_{\ge}
={1\over 2\pi i}\oint_C dz
{\left(1-{1\over z^2}\right) Z_4^{z,w}(s,x,z^2 ) \over z-\sqrt{y}}
\eea
The contour $C$ must again have a radius larger than $\sqrt{y}$, so we again choose the unit
circle $|z|=1$. The integrand has poles at 
$z=\pm s \sqrt{x}$, $z=\sqrt{y}$, $z=\pm \frac{1}{s \sqrt{x}}$, $z=\pm i sx$, 
$z=-\frac{s \sqrt{x}}{2} i \left(1\pm i\sqrt{3} \right)$, $z=\pm \frac{i}{sx}$ and
$z=-\frac{\left(1\pm i\sqrt{3}\right)}{2 s \sqrt{x}}$.
The integral above receives contributions from the poles at $z=\pm s \sqrt{x}$, $z=\sqrt{y}$, $z=\pm i sx$,
and $z=-\frac{s \sqrt{x}}{2} i \left(1\pm i\sqrt{3} \right)$.
We obtain
\bea\label{G4form} 
G_4^{z,w}(s,x,y) 
={s^4 R(s,x,y)\over 
(1-s^2 x y)(1-s^3 x^{3\over 2} y^{3\over 2})(1-s^4 x^2 y^2)(1-s^4 x^2)\left(1-s^6 x^3\right)
\left(1-s^8 x^4\right)}\cr
\eea
where
\bea
R(s,x,y)&=&1+s^5 x^{5\over 2} \big(\sqrt{y}+s^3 x^{3\over 2} y+s^5 x^{5\over 2} y+y^3-s^6 x^3 y^{5\over 2}
-s^8 x^4 y^{5\over 2}-s^{16} x^8 y^{7\over 2}-s^{11} x^{11\over 2} y^2 \left(1+y\right)\cr
&&+s^7 x^{7\over 2} \left(1-y^2\right)+s^4 x^2 y^{3\over 2} \left(1-y^2\right)+s^2 x\sqrt{y} \left(1+y^2\right)
-s^9 x^{9\over 2} y \left(1+y^2\right)\cr
&&-s^{10} x^5 y^{3\over 2} \left(1+y-y^2\right)-s \sqrt{x} \left(1-y-y^2\right)\big)
\eea
It is again easy to check, using mathematica, that this expression does indeed have the correct expansion.

There are other sectors of primaries that are slight variations of  the extremal sector studied in this section.  Polynomials in $z_I,\bar w_I $ correspond to primaries with $ ( \Delta = n + q , J_3^R = q )$. Polynomials in $\bar z_I,w_I$ 
correspond to primaries with $ ( \Delta = n + q , J_3^R = - q )$. Polynomials in $\bar z_I,\bar w_I$
 correspond to primaries with $  ( \Delta = n + q , J_3^L = - q )$. 

\section{\label{sec:Construction} Construction and construction with symmetric groups}

In this section we would like to provide construction formulas for the extremal primaries we have counted in 
section \ref{sec:Counting}.
To accomplish this the polynomial representation of $SO(4,2)$ introduced in section \ref{sec:PolyRep} will play
a central role.
These polynomials are constructed using the coordinates $x^I_\mu$, $I=1,...,n$ which admit a natural action of $S_n$.
Constructing primaries then amounts to constructing polynomials that are consistent with (\ref{constraints}).
The first of (\ref{constraints}) can be satisfied by constructing $n-1$ translationaly invariant ``relative coordinates'' 
out of the $x^I_\mu$.
This construction is not unique.
Following \cite{BHR2}, a particularly convenient choice makes use of the variables
\begin{eqnarray}
X_\mu^{(a)}={1\over\sqrt{a(a+1)}}(x^1_\mu+\cdots+x^a_\mu-ax^{a+1}_\mu)\label{hookvars}
\end{eqnarray}
These variables are in the $[n-1,1]$ irrep of $S_n$.
To satisfy the second of (\ref{constraints}) we need to build polynomials that are harmonic.
In terms of complex coordinates the Laplacian is
\bea
\sum_\mu {\partial\over\partial x^I_\mu}{\partial\over\partial x^I_\mu}
={\partial\over\partial z^I}{\partial\over\partial \bar z^I}
+{\partial\over\partial w^I}{\partial\over\partial \bar w^I}
\eea
It is clear that we can build harmonic polynomials by considering polynomials that are functions only of the 
$z^I$, which gives the leading twist primaries, or that are functions of the $z^I$ and $w^I$, which gives the
leading left twist primaries considered in section \ref{sec:LS}.
The harmonic constraint is the only constraint that is not first order. 
By replacing this with a holomorphic constraint, which is first order, the resulting problem entails finding families
of polynomials that obey first order equations.
This implies that the problem will now have a natural ring structure, something which will be visible in our construction.
The final constraint that needs to be obeyed is that the polynomials are $S_n$ invariants.
The counting formulas we derived in the previous section will give valuable insight into how to handle this final 
constraint.

\subsection{Leading Twist Primaries}

Specializing to $n=3$ and employing complex variables, we have
\begin{eqnarray}
Z^{(1)}={z^1-z^2\over\sqrt{2}}\qquad Z^{(2)}={z^1+z^2-2z^3\over\sqrt{6}}
\end{eqnarray}
plus the obvious formulas for $\bar Z^{(a)}$, $W^{(a)}$ and $\bar W^{(a)}$.
The nice thing about these variables is that $S_n$ acts on these variables with Young's orthogonal 
representation of $[n-1, 1]$, i.e. for $n=3$ we have\cite{Hammermesh}
\begin{eqnarray}
\Gamma_{\tiny\yng(2,1)}\left(\, (12)\, \right)=\left[
\begin{matrix}
-1 &0\cr
0 &1
\end{matrix}\right]
\qquad
\Gamma_{\tiny\yng(2,1)}\left(\, (23)\,\right)=\left[
\begin{matrix}
{1\over 2} &{\sqrt{3}\over 2}\cr
{\sqrt{3}\over 2} &-{1\over 2}
\end{matrix}\right]\nonumber
\end{eqnarray}
The remaining elements of the group can be generated using these two.
When acting on a product of variables, say $Z^{(a_1)}Z^{(a_2)}\cdots Z^{(a_k)}$ we have
\begin{eqnarray}
  \Gamma_k (\sigma)=\Gamma_{\tiny\yng(2,1)}(\sigma)\times\cdots\times\Gamma_{\tiny\yng(2,1)}(\sigma)
\end{eqnarray}
where we take a tensor product (the usual Kronecker product) of $k$ copies of the matrices of the hook irrep.
Any polynomial in the hook variables automatically obeys (\ref{constraints}).
Thus, all that is left is to project to $S_n$ invariants in $V_{H}^{\otimes k}$.
We can build these by acting with the projector from the tensor product of $k$ copies of the hook
onto the trivial irrep
\begin{eqnarray}
   P_{{\tiny \yng(3)}}={1\over 3!}\sum_{\sigma\in S_3}\Gamma_k (\sigma)\label{IdProj}
\end{eqnarray}
Acting on $Z^{\otimes k}$ we obtain an expression of the form $\sum_i \hat{n}_i P_i(z)$ where $\hat{n}_i$ are 
unit vectors inside the carrier space of $\,\,{\tiny \yng(2,1)}^{\otimes k}$ and $P_i(z)$ are the polynomials that
can be translated into primary operators.

It is useful to consider a few examples.
Acting with the projector (\ref{IdProj}) on the tensor product of $k$ copies of the hook, we find
\bea
   P_{a_1 a_2\cdots a_k}=\sum_{\sigma\in S_3}\Gamma_k (\sigma)_{a_1 a_2\cdots a_k,b_1 b_2\cdots b_k}
Z^{(b_1)}Z^{(b_2)}\cdots Z^{(b_k)}
\eea
It is simple to implement this projector in mathematica.
For $k=1$ we find $P_{a_1}=0$. 
For $k=2$ the projector is
\bea
P_{a_1 a_2}=((z_1 - z_2)^2 + (z_1 - z_3)^2 + (z_2 - z_3)^2)\left[
\begin{matrix}
{1\over 6} \cr 0\cr 0\cr {1\over 6}
\end{matrix}
\right]
\eea
so the invariant polynomial is
\bea
  P(z)=((z_1 - z_2)^2 + (z_1 - z_3)^2 + (z_2 - z_3)^2)
\eea
By inspection, this obviously obeys (\ref{constraints}).
For $k=3$ the projector is
\bea
P_{a_1 a_2 a_3}=
((z_1 + z_2 - 2 z_3) (z_1 + z_3 - 2 z_2) (z_2 + z_3 - 2 z_1))\left[
\begin{matrix}
0 \cr {-1\over 6\sqrt{6}} \cr {-1\over 6\sqrt{6}} \cr 0 \cr {-1\over 6\sqrt{6}} \cr 0 \cr 0 \cr {1\over 6\sqrt{6}}
\end{matrix}
\right]
\eea
so the invariant polynomial is
\bea
  P(z)=((z_1 + z_2 - 2 z_3) (z_1 + z_3 - 2 z_2) (z_2 + z_3 - 2 z_1))
\eea
This polynomial again obeys (\ref{constraints}).
Finally, for $k=4$ the projector is
\bea
P_{a_1 a_2 a_3 a_4}=
((z_1 - z_2)^4 + (z_1 - z_3)^4 + (z_2 - z_3)^4)
 \left[
\begin{matrix}
{1\over 12}\cr 0\cr 0\cr {1\over 36}\cr 0\cr {1\over 36}\cr {1\over 36}\cr 0\cr 0\cr {1\over 36}\cr {1\over 36}\cr 
0\cr {1\over 36}\cr 0\cr  0\cr {1\over 12}
\end{matrix}
\right]
\eea
so the invariant polynomial is
$$
  P(z)=((z_1 - z_2)^4 + (z_1 - z_3)^4 + (z_2 - z_3)^4)
$$
This clearly obeys (\ref{constraints}), so this is again the correct answer.

The polynomials we construct in this way will obey the conditions spelled out in (\ref{constraints}).
In fact, they obey an even stronger linear condition
\bea
\partial_{\bar z^I}P(\vec z)=0=\partial_{\bar w^I}P(\vec z)
\eea
which imply the Laplacian constraint.
As a result, taking all possible values of $k$ we find that the polynomials constructed in this way exhibit a 
highly non-trivial structure enjoyed by the leading twist primaries: the polynomials
$P_i(z)$ are a finitely generated polynomial ring.
The counting formula (\ref{Polyinz}) gives the Hilbert series for holomorphic functions on
$(\mC^n/\mC)/S_n$. The quotient by $\mC$ sets the center of mass momentum of the many body
wave function to zero as dictated by the first of (\ref{constraints}).
The orbifold by $S_n$ implements the last of (\ref{constraints}). 
The counting formula (\ref{Polyinz}) implies that the ring has $n-1$ generators.
These generators are given by constructing the $n-1$ possible independent $S_n$ invariants out of the hook
variables introduced in (\ref{hookvars}). 
For example, for $n=2$ fields the polynomials are generated by $(z_1-z_2)^2$.
The polynomials corresponding to primaries are
\begin{eqnarray}
(z_1-z_2)^{2k}
\end{eqnarray}
Using (\ref{polytranslate}) it is easy to see that (these vanish if $s$ is odd)
\begin{eqnarray}
O_s &=& (z_1-z_2)^s\cr
&\leftrightarrow&{s!\over 2^s}\sum_{k=0}^s 
{(-1)^k \over (k! (s-k)!)^2}\, \partial_z^{s-k}\phi\, \partial_z^{k}\phi
\end{eqnarray}
reproducing the higher spin currents, given for example in\cite{Giombi:2016hkj}.
For $n=3$ fields the ring of polynomials that correspond to primary operators is generated by
\begin{eqnarray}
(z_1-z_2)^{2}+(z_1-z_3)^{2}+(z_2-z_3)^{2}
\end{eqnarray}
and
\begin{eqnarray}
(z_1+z_2-2z_3) (z_3+z_2-2z_1) (z_1+z_3-2z_2)
\end{eqnarray}
In general, the generators of the ring are a product of the variables $Z^{(a)}$ introduced above, such that the product
is $S_n$ invariant. For $n=4$ the ring is generated by $(z_1-z_2)(z_2-z_1)+\cdots$, 
$(z_1+z_2-2z_3) (z_3+z_2-2z_1) (z_1+z_3-2z_2)+\cdots$ and
$(z_1+z_2+z_3-3z_4) (z_3+z_2+z_4-3z_1) (z_1+z_3+z_4-3z_2)(z_1+z_2+z_4-3z_3)$, where $\cdots$ stand for
terms that must be summed to obtain an $S_4$ invariant.
The ring structure that has appeared is rather interesting.
The product on the ring is simply multiplication of polynomials.
This is a natural product in the polynomial language, but is highly non-trivial in the original CFT description.
A natural guess would be that this is somehow connected to the OPE of primaries, which is the natural product
on the primaries of the CFT.
However, this cannot  be correct because the polynomial ring exists for a fixed number $n$.
Thus, in terms of the CFT language, the ring multiplication 
 is  a product between two primaries, each of which has $n$ fields, and the
result is again a primary with $n$ fields.
The operator product of two local operators, each containing $n$ fields, is a sum of operators containing 
$2n - 2k$ fields with $k=0,1,...,n$.
For odd $n$ the product of elements of the ring gives an operator with $n$ fields. 
This product  can therefore not  even be a subalgebra of the CFT operator product algebra. 
This product and the associated ring  structure of primary fields in free CFT4 
appears to be a genuinely new structure, not previously noticed. 

A natural question to ask is whether or not these primary operators are orthogonal.
We can translate any polynomial into an operator and then compute the two point function of the operator.
The computation can also be carried out by a judicious choice of an inner product for the polynomial.
For example, consider the correlator
\begin{eqnarray}
\langle \partial_{z}^{k}\phi (x)\partial_{z'}^{l}\phi (x')\rangle
= (-1)^k(k+l)!
{(\bar z-\bar z')^{k+l}\over (|z-z'|^2+|w-w'|^2)^{k+l+1}}
\end{eqnarray}
Everything in the above result is determined by conformal invariance, except the overall number $=(-1)^k (k+l)!$.
Recalling that $z^n$ translates into ${1\over n!}\partial_z^n$, this number can be computed if we use the following
inner product for the polynomials
\begin{eqnarray}
\langle z^k z^{\prime l}\rangle_p =(-1)^k {(k+l)!\over k!l!}
\end{eqnarray}
Notice that the norm following from this inner product is not positive definite.
For $n$ fields we have polynomials in $z_{k}$ for the primary at $x$ and in $z'_k$ for the primary at $x'$,
with $k=1,...,n$.
In this more general setting, the inner product is
\begin{eqnarray}
\langle \prod_{k=1}^n z_k{}^{p_k} \prod_{l=1}^n z'_l{}^{q_l}\rangle_p =
\prod_{k=1}^n (-1)^{p_k} {(p_k+q_k)!\over p_k!q_k!}
\end{eqnarray}
In addition, due to Wick's theorem, there are a total of $n!$ Wick contractions contributing, which introduces a 
factor of $n!$.
In the end, if polynomials $P_i$ of degree $k_i$ in $n$ variables translate into primaries ${\cal O}_i$ constructed
from $n$ fields with dimension $n+k_i$, then we have
\begin{eqnarray}
\langle {\cal O}_i (x) {\cal O}_j (x')\rangle=
{c_{ij}(\bar z-\bar z')^{k_i+k_j}\over (|z-z'|^2+|w-w'|^2)^{k_i+k_j+n}}
\end{eqnarray}
with
\begin{eqnarray}
c_{ij}=n!\langle P_i(z_k) P_j(z'_k)\rangle_p
\end{eqnarray}
Using the above formulas, it is easy to check that primary operators with different dimensions are orthogonal, as they
must be.
Further, we also see that although our ring of primaries is a basis, the operators in the basis are not orthogonal.

\subsection{Extremal Primaries}

The above construction is easily extended to the other classes of extremal primaries we have counted.
The leading left or right twist class is provided by polynomials in two holomorphic coordinates, $z$ and $w$.
Consider polynomials of degree $k$ in $Z$ and of degree $l$ in $W$, with $Z,W$ the hook variables transforming 
in the hook representation $V_H$ of $S_n$, described by a Young diagram with row lengths $[n-1,1]$. 
These polynomials belong to a subspace of $V_H^{\otimes k}\otimes V_H^{\otimes l}$ of $S_n$.
To characterize this subspace using representation theory, start with the decompositions in terms of $S_n\times S_k$
irreps
\begin{eqnarray}\label{decompsSnSkSl} 
&& V_H^{ \otimes k } = \bigoplus_{ \Lambda_1 \vdash n , \Lambda_2 \vdash k } V_{ \Lambda_1}^{  (S_n)} \otimes V_{ \Lambda_2}^{ (S_{k} ) } \otimes V^{ Com ( S_n \times S_k )}_{ \Lambda_1 , \Lambda_2} \cr 
&& V_H^{ \otimes l  } = \bigoplus_{ \Lambda_3 \vdash n , \Lambda_4 \vdash l  }  V_{ \Lambda_3}^{  (S_n)} \otimes V_{ \Lambda_4}^{ (S_{l} ) } \otimes V^{ Com ( S_n \times S_l )}_{ \Lambda_3 , \Lambda_4} 
\end{eqnarray}
$ Com  ( S_n \times S_k ) $ is the algebra of linear operators on $V_{H}^{\otimes k }$ 
which commute with $ S_n \times S_k$. 
The tensor product   $ V_H^{ \otimes k } \otimes V_{H}^{ \otimes l }$ 
is a representation of 
\bea\label{theAlgebra}  
\mC ( S_n ) \otimes \mC ( S_k ) \otimes \mC ( S_n ) \otimes \mC ( S_l ) 
\eea
 These decompositions (\ref{decompsSnSkSl})  have been studied in detail in \cite{BHR2} where they were used to construct BPS states
of ${\cal N}=4$ SYM. 
In the application we consider here, the $Z$ and $W$ variables are commuting which implies that they are in the 
trivial rep $\Lambda_2\otimes\Lambda_4=[k]\otimes [l]$ of $S_k\times S_l$.
The multiplicity with which a given $S_n\times S_k$ irrep $(\Lambda_1,\Lambda_2)$ appears is given by the
dimension of the irrep of the commutants $Com ( S_n \times S_l )$ in $V_H^{\otimes k}$.
We want to project to states in $V_{H}^{\otimes k}\otimes V_H^{\otimes l}$ which are invariant under the diagonal $\mC ( S_n ) $ in the algebra (\ref{theAlgebra}). 
This constrains $\Lambda_3=\Lambda_1$. 
Thus we find that the number of $ S_k \times S_l \times S_n$ invariants is 
\begin{eqnarray}
\sum_{\Lambda_1\vdash n} {\rm Mult}(\Lambda_1,[k];S_n\times S_k) ~~ {\rm Mult}(\Lambda_1,[l];S_n\times S_l) 
\end{eqnarray}
The generating functions for these multiplicities have been derived in \cite{BHR2}. 
$Mult (\Lambda_1,[k];S_n\times S_k)$ is the coefficient of $q^k$ in 
\begin{eqnarray}
Z_{SH} (q;\Lambda_1)&=&( 1- q ) ~  q^{\sum_{i} c_i(c_i-1)\over 2} ~ \prod_b{1\over (1-q^{h_b})}\cr
&=&\sum_{k}q^k Z_{SH}^k (\Lambda_1) \label{Mult}
\end{eqnarray}
Here $c_i$ is the length of the $i$'th column in $ \Lambda_1$, $b$ runs over boxes in the Young diagram $\Lambda_1$ 
and $h_b$ is the hook length of the box $b$. 
Thus, for the number of primaries constructed from $ z_i , w_i $ we get 
\begin{eqnarray} 
\sum_{ \Lambda_1 \vdash n } Z_{SH}^k  ( \Lambda_1 )
Z_{SH}^l ( \Lambda_1 )
\end{eqnarray}
The above integer gives the number of primaries in the free scalar theory, of weight $n+k+l$, with spin
$(J_3^L,J_3^R)=({k+l\over 2},{k-l\over 2})$.
For the generating function  $ Z^{z,w}_n(s,x,y)$  which encodes all $ k , l $, we have 
 \bea\label{genZzw} 
   Z^{z,w}_n(s,x,y)=s^n\sum_{\Lambda_1\vdash n}Z_{SH}(s\sqrt{xy},\Lambda_1)Z_{SH}(s\sqrt{x\over y},\Lambda_1)
 \eea
where $\Lambda_1$ is a partition of $n$ and we can use the formula \ref{Mult}. 

We can in fact see  that the above discussion is  consistent with the Taylor expansion formula (\ref{TaylExpForm}). 
We can recognise this formula as ${\rm Tr}(P_{[n]}a^{L_0}b^{L_0}) $ where the trace is being taken in 
\bea 
\bigoplus_{ k , l = 0 }^{ \infty }  \Sym^{ k + l }  ( V_H  )
\eea
which can be identified with a tensor product of discrete irreps of $SL(2)$, which we may denote as  
 $ V_{ SL(2) }^{\otimes n }   \otimes V_{ SL(2)}^{ \otimes n }  $: one factor corresponds to 
 the $z$ variables and another to the $w$ variables.  $P_{[n]} $ is the projector for the symmetric irrep of $S_n$. 
 Factor out the trace into the separate $ SL(2) $ factors to get (see (\ref{forzzw}))
\bea
{1\over n!}{\partial^n F_2\over\partial t^n}\Big|_{t=0} &=& 
{\rm Tr}_(P_{[n]}a^{L_0}b^{L_0})\cr
&=&\sum_{\Lambda_1 \vdash n }{\rm Tr } \left ( P_{\Lambda_1} a^{L_0}) {\rm Tr} (P_{\Lambda_1}b^{L_0}\right ) 
\eea
Note also that
 \bea
 { 1 \over ( 1- a )  } Z_{SH}(a,\Lambda)={\Tr}(P_{\Lambda }a^{L_0})
 \eea
which follows by recognising that the raising operators of the $SL(2)$ representation on 
$z_1 \cdots z_n$ can be separated into a weight one centre of mass coordinate 
and the differences which span the hook representation of $S_n$. 
This demonstrates the equivalence between the Taylor expansion formula (\ref{TaylExpForm}) 
 and the $ S_n \times S_k \times S_l $ formula 
(\ref{genZzw}). It is important to note that this is a non-trivial equivalence: both formulae are self-contained ways of calculating 
the multiplicities.

We have thus  re-expressed  our earlier Taylor expansion  in a way that makes the representation theory content of the
counting manifest. This structure in the counting problem can be used to provide an explicit construction formula.
First we need to decompose the $Z$ and $W$ polynomials into definite $S_n$ irreps.
The projector onto irrep $r$ from the tensor product of $k$ copies of the hook is
\begin{eqnarray}
 P^{r,k}={1\over n!}\sum_{\sigma\in S_n}\chi_r(\sigma)\Gamma_k (\sigma)
\end{eqnarray}
We also need the projection onto the symmetric irrep
\begin{eqnarray}
P^{k+l}={1\over n!}\sum_{\sigma\in S_n} \Gamma_{k+l} (\sigma)
\end{eqnarray}
Using these two projectors, the polynomials corresponding to primaries constructed using two holomorphic
variables are now given by
\begin{eqnarray}
\sum_A P_A(z,w)\hat n^A
=P^{k+l}\,\sum_{r\vdash n} (P^{r,l}\times P^{r,k}) Z^{\otimes k}W^{\otimes l}\label{zwconst}
\end{eqnarray}
where $\hat{n}^A$ are unit vectors inside the carrier space of  $ {\tiny \yng(2,1)}^{\otimes \, k+l}$ and $P_A(z,w)$ 
are the polynomials we want.
In fact, the construction formula given in (\ref{zwconst}) constructs a larger class of polynomials than those
counted in  (\ref{zwcount}).
This is because the polynomials counted in (\ref{zwcount}) are extremal and hence they are annihilated by 
$J_+^R$.
We will return to this point in the discussion below.

The construction formula that has been sketched above can easily be implemented numerically.
To implement (\ref{zwconst}),  we need the projector onto irrep $r$ in the space obtained by taking 
the tensor product of $k$ copies of the hook
\bea
 P^r_{a_1 \cdots a_k,b_1 \cdots b_k}={1\over n!}\sum_{\sigma\in S_n}\chi_r(\sigma)
\Gamma_k (\sigma)_{a_1 a_2\cdots a_k,b_1 b_2\cdots b_k}
\eea
and we need the projection onto the symmetric irrep
\bea
P_{a_1\cdots a_n,b_1\cdots b_n}={1\over n!}\sum_{\sigma\in S_n}
\Gamma_k (\sigma)_{a_1 \cdots a_n\, ,\, b_1 \cdots b_n}
\eea
We find that (\ref{zwconst}) is now given by
\bea
\sum_A P_A(z,w)\hat n^A_{e_1\cdots e_{k+l}}
&=&P_{e_1\cdots e_{k+l},a_1\cdots a_k c_1\cdots c_l}
P^r_{a_1 \cdots a_k,b_1 \cdots b_k}P^r_{c_1 \cdots c_l,d_1 \cdots d_l}
Z^{(b_1)}\cdots Z^{(b_k)}W^{(d_1)}\cdots W^{(d_l)}\cr
&=&\tilde P_{e_1\cdots e_{k+l}}
\nonumber
\eea
where $\hat{n}^A$ are unit vectors and $P_A(,wz)$ are the polynomials we want.
To start, consider $k=l=1$. 
We find
\bea
\tilde P_{e_1 e_2}=
(-w_3 (z_1 + z_2 - 2 z_3) + w_1 (2 z_1 - z_2 - z_3) - w_2 (z_1 - 2 z_2 + z_3))
\left[
\begin{matrix}
1 \cr 0 \cr 0 \cr 1
\end{matrix}
\right]
\eea
so that the invariant polynomial is
\bea
P(z,w)=-w_3 (z_1 + z_2 - 2 z_3) + w_1 (2 z_1 - z_2 - z_3) - w_2 (z_1 - 2 z_2 + z_3)
\eea
This polynomial is not extremal, which is easily verified by computing
\bea
-J_+^R P(z,w)=z_i{\partial\over\partial w_i}P(z,w)=(z_1 - z_2)^2 + (z_1 - z_3)^2 + (z_2 - z_3)^2
\eea
so that this is another state in the multiplet of the $k=2$ primary we built in the last section.

To focus on the extremal polynomials counted in (\ref{zwcount}) we must implement the constraint that these
polynomials are annihilated by $J_+^R$.
Towards this end, note that the polynomials in $Z,W$ carry a representation of $SU(2)_R$, so that we can further
decompose the polynomials according to their $SU(2)_R$ quantum numbers.
$Z,W$ form an $SU(2)$ doublet with $Z$ the $+{1\over 2}$ state and $W$ the $-{1\over 2}$ state.
There is an action of $S_{k+l}$ on these polynomials that commutes with $SU(2)_R$.
This $S_{k+l}$ action acts to permute the $W^{(a)}$ and $Z^{(b)}$ factors.
Denote the matrix representing $\sigma\in S_{k+l}$ by $\Gamma (\sigma)$. 
This rep is generated by the adjacent permutations which are easy to build. 
Towards this end, note that swapping two factors in the tensor product is accomplished by
the permutation $P$ which obeys $P\, x\otimes y=y\otimes x$, i.e. we have
\bea
P\left[\begin{matrix} x_1 y_1\cr x_1 y_2\cr x_2 y_1\cr x_2 y_2\end{matrix}\right]
=\left[\begin{matrix} x_1 y_1\cr x_2 y_1\cr x_1 y_2\cr x_2 y_2\end{matrix}\right]\quad\Rightarrow\quad
P=\left[\begin{matrix} 1 &0 &0 &0\cr 0 &0 &1 &0\cr 0 &1 &0 &0\cr 0 &0 &0 &1\end{matrix}\right]
\eea
Using the adjacent permutations we can construct any $\Gamma (\sigma)$ and then any projector
\bea
K^R={1\over (k+l)!}\sum_{\sigma\in S_{k+l}}\chi_R(\sigma)\Gamma (\sigma)
\eea
with $\chi_R(\sigma)$ a symmetric group character.
The label $R$ is a Young diagram with at most 2 rows.
The spin of the $SU(2)$ irrep that $K^R$ projects to is given by $(R_1-R_2)/2$ where $R_1$ and $R_2$ are the
lengths of the rows of $R$.
As an example, consider $k=2=l$. 
The rep of $S_4$ we need is generated by (${\bf 1}$ is the $2\times 2$ identity)
\bea
\Gamma ((12))=P\otimes {\bf 1}\otimes {\bf 1} \qquad
\Gamma ((23))={\bf 1}\otimes P\otimes {\bf 1} \qquad
\Gamma ((34))={\bf 1}\otimes {\bf 1}\otimes P 
\eea
To construct the primary corresponding to $s^7 x^2$ we need to project on the $SU(2)_R$ irrep with spin zero.
This is accomplished by using the projector
\bea
K^{\tiny\yng(2,2)}_{a_1 a_2 a_3 a_4,b_1 b_2 b_3 b_4}={1\over 4!}\sum_{\sigma\in S_4}\chi_{\tiny \yng(2,2)}(\sigma)
\Gamma_{a_1 a_2 a_3 a_4,b_1 b_2 b_3 b_4}(\sigma)
\eea
It is simple to compute
\bea
K^{\tiny\yng(2,2)}_{a_1 a_2 a_3 a_4,b_1 b_2 b_3 b_4} \tilde P_{b_1 b_2 b_3 b_4}
=(w_1 (z_2 - z_3) + w_2 (z_3 - z_1) + w_3 (z_1 - z_2))^2
 \left[
\begin{matrix}
0\cr 0\cr 0\cr {4\over 3}\cr 0\cr -{2\over 3}\cr -{2\over 3}\cr 0\cr 0\cr -{2\over 3}\cr -{2\over 3}\cr 0\cr {4\over 3}\cr 0\cr  0\cr 0
\end{matrix}
\right]
\eea 
Thus the invariant polynomial is
\bea
P(z,w)=(w_1 (z_2 - z_3) + w_2 (z_3 - z_1) + w_3 (z_1 - z_2))^2
\eea
By inspection it is obvious that this polynomial obeys the conditions (\ref{constraints}) and further that it is
a highest weight of $SU(2)_R$, i.e. $J_+^R  P(z,w)=0$.
The above polynomial suggests a natural generalization: consider the family of polynomials indexed by
the integer $n$
\bea
\Psi_n=\Big( w^{(3)} (\bar z^{(2)}-\bar z^{(1)}) + w^{(2)} (\bar z^{(1)} - \bar z^{(3)}) + w^{(1)} 
(\bar z^{(3)}-\bar z^{(2)})\Big)^{2n}
\eea
It is obvious that they also obey (\ref{constraints}) and hence that these polynomials do correspond to primary
operators.
It is also clear that they are extremal, i.e. $J_+\Psi_n=0$.
These primaries have spin $[2n,0]$ and dimension $\Delta =3+4n$.
The translation into the free field language is
\bea
{\cal O}^{\Delta=4n+3}_{[2n,0]}
=\sum_{r=0}^{2n}\sum_{s=0}^{2n-r}\sum_{t=0}^r\sum_{u=0}^s\sum_{v=0}^{2n-r-s}
{(2n)!(-1)^{t+u+v}\over (r-t)!t!(s-u)!u!(2n-r-s-v)!v!}\times\cr
\cr
\times (P_w^{2n-r-s}P_{\bar z}^{t+s-u}\phi)
(P_w^{s}P_{\bar z}^{r-t+v}\phi)
(P_w^{r}P_{\bar z}^{2n+u-r-s-v}\phi)
\eea

The polynomials we have constructed in (\ref{zwconst}) obey all of the conditions spelled out in (\ref{constraints}).
In fact, they again obey an even stronger linear condition
\bea
\partial_{\bar z^I}P(\vec z,\vec w)=0=\partial_{\bar w^I}P(\vec z,\vec w)
\eea
which imply the Laplacian constraint.
As a result, taking all possible values of $k,l$ we find that the polynomials $P_A(z,w)$ are again a finitely 
generated polynomial ring.
This is a consequence of the Leibnitz rule for the derivatives of a product of functions.
The ring of polynomials that correspond to extremal primaries is the polynomial ring of holomorphic functions for
\bea\label{CYorbs}  
 {  (\mC^2 )^n / (  \mC^2 \times S_n ) }
\eea

In (\ref{nis3zw}), we have computed the Hilbert series for the polynomials in two holomorphic variables, that 
correspond to extremal primary operators built using two scalar fields.
Using generalities about Hilbert series for algebraic varieties (see \cite{GeomAP,SQCDAd} 
for applications in the context of moduli spaces of SUSY gauge theories), we know that if the ring is generated by $h$
homogeneous elements of positive degrees $d_1,\cdots ,d_h$, then the Hilbert series is a rational fraction
\bea
HS(t)=\frac {Q(t)}{\prod _{i=1}^{h}(1-t^{d_{i}})}
\eea
where Q is a polynomial with integer coefficients.
Thus, we see from (\ref{nis3zw}) that for $n=3$ the polynomials $P_A(z,w)$ are a finitely generated polynomial 
ring with $4$ generators and one relation and that this space of polynomials is a complete intersection and it is 3
dimensional.
Using this Hilbert series and the explicit constructions described above, we can identify the generators 
($z_{ij}\equiv z_i-z_j$)
\begin{eqnarray}
G_1=(z_{12})^{2}+(z_{13})^{2}+(z_{23})^{2} &\leftrightarrow& s^2 xy\cr\cr
G_2=(z_{13}+z_{23}) (z_{31}+z_{21}) (z_{12}+z_{32}) &\leftrightarrow& s^3\sqrt{x^3y^3}\cr\cr
G_3=\left|
\begin{matrix}
w_1 &w_2 &w_3\cr z_1 &z_2 &z_3\cr 1 &1 &1
\end{matrix}\right|^{2}
&\leftrightarrow&  s^4 x^2\cr\cr
G_4=\left|
\begin{matrix}
z_1^2 &z_2^2 &z_3^2\cr z_1 &z_2 &z_3\cr 1 &1 &1
\end{matrix}
\right|\times\left|
\begin{matrix}
w_1 &w_2 &w_3\cr z_1 &z_2 &z_3\cr 1 &1 &1
\end{matrix}
\right|
&\leftrightarrow&  s^5 x^{5\over 2}y^{3\over 2}
\end{eqnarray}
of this ring.
Consider the last generator above: either of the determinants being multiplied is antisymmetric under
permuting $1,2$ or $1,3$ or $2,3$ so that the product is symmetric.
The relation obeyed by these generators is easily identified
\bea
27 \left( G_4 \right)^2 + G_3 \left((G_2)^2-{1\over 2}(G_1)^3\right)=0
\eea   
Once again the ring structure exhibited by the polynomials implies a genuinely new structure for the extremal
primary operators that was not previously recognized.
The Hilbert series in more complicated situations encodes detailed information about the generators of the ring,
relations between these generators, relations between the relations and so on.
An example of this structure is given in Appendix \ref{HS}.

The Hilbert series we have computed so far exhibit a palindromic property of the numerators.
This can be verified for $ Z^{\rm z,w}_3 ( s, x, y)$ and $Z^{\rm z,w}_4 ( s, x, y )$. 
A general property of the numerators
\bea 
Q_n ( s , x , y )   = \sum_{ k =0 }^D a_k ( x, y ) s^k 
\eea
is that $ a_{ D-k}  (  x , y ) = a_{ k} ( x , y  )$.
A theorem due to Stanley\cite{stanley} suggests that this palindromic property of the numerators implies the Calabi-Yau
property of the underlying orbifolds.  
It is fascinating that non-trivial properties of  the combinatorics of primary fields in four dimensional scalar field theory 
is related to the  geometry of Calabi-Yau orbifolds (\ref{CYorbs}). 
Motivated by this connection, we will prove this palindromic property of the numerators in the next section.
 
To obtain $G^{z,w}_n(s,x,y)$ from $Z^{z,w}_n(s,x,y)$, we have kept only the highest weight operator 
(under $SU(2)$) from a complete
spin multiplet of primary operators.
Geometrically, this can be viewed as modding out by the action of $G_+$, generated by the $SU(2)$ raising operator
$J^+$, i.e. $G_+$ is the unipotent group of upper triangular $2\times 2$ matrices with 1 on the diagonal.
Consequently, the Hilbert series $ G_n ( s , x, y ) $ is the polynomial ring of functions for
\bea 
 { ( \mC^2 )^n   \over  ( \mC^2 \times  G_+ \times S_n )   } 
\eea

 \subsection{ Palindromy properties } 

The palindromic property of the Hilbert series can be stated as follows
\bea
Z^{z,w}_n ( q_1^{-1} , q_2^{-1}  )  = ( q_1 q_2)^{ n-1 }Z^{z,w}_n ( q_1 , q_2 )
\eea
In this section we will prove that our Hilbert series $Z^{z,w}_n (q_1,q_2)$ do indeed enjoy this transformation
property.

Our starting point is the formula
\bea 
 Z^{z,w}_n ( q_1 , q_2 ) = s^n \sum_{ \Lambda \vdash n } Z_{SH}  ( q_1  , \Lambda ) Z_{ SH} ( q_2 , \Lambda )  
\eea
where $ q_1 = s \sqrt{ xy } , q_2 = s \sqrt{ x /y} $. 
This has the property $ Z_{n}^{z,w} ( q_1 , q_2 ) = Z_n^{z,w} ( q_2 , q_1 )$. 
The exchange of $q_1 , q_2 $ amounts to the inversion of $y$. 
Now, observe that 
\bea
Z_{SH} ( q^{-1} , \Lambda )  = (-q)^{ n-1} Z_{SH} ( q , \Lambda^T ) \label{PalZ}
\eea
This is easily demonstrated using the explicit formula (\ref{Mult}) and the identity
\bea 
\sum_b h_b & = &  { 1 \over 2 } \left ( \sum_{ i } c_i ( c_i +1 )  + \sum_i r_i ( r_i +1) \right ) - n \cr  
    & = & { 1 \over 2  } \left ( \sum_i c_i^2 - \sum_i r_i^2  \right ) 
\eea
Here $c_i$ is the length of the $i$'th column and $r_i$ is the length of the $i$'th row. 
Also note that the row lengths of $ \Lambda^T$ are the column lengths of $ \Lambda $ and vice versa. 
The identity can be understood as follows. As we sum over hook lengths, for each column of length 
$c_i$ we have a contribution to the sum  of $ 1 + 2 + \cdots + c_i $ as we start from the bottom and go up 
to the top. For each row, we can similarly sum $ 1+2 \cdots + r_i$, but this over counts $1$ for each box. 
Hence the identity. 
Using this result
\bea 
Z^{z,w}_n ( q_1^{-1} , q_2^{-1}  ) & = &  s^n
( q_1 q_2)^{ n-1 }  \sum_{ \Lambda \vdash n } Z_{SH}  ( q_1  , \Lambda^T  ) Z_{ SH} ( q_2 , \Lambda^T  ) \cr 
& = & s^n ( q_1 q_2)^{ n-1 }  \sum_{ \Lambda \vdash n } Z_{SH}  ( q_1  , \Lambda ) Z_{ SH} ( q_2 , \Lambda  )\cr
& = &  ( q_1 q_2)^{ n-1 }Z^{z,w}_n ( q_1 , q_2 )\label{palin}
\eea
In the last step, we used the fact that transposition is a symmetry of the set of Young diagrams.
Summing over $ \Lambda^T $ is the same as summing over $ \Lambda$. 

The Hilbert series $G_n^{z,w}(s,x,y)$ also exhibit the palindromy property.
We know
\bea
   Z^{z,w}_n (s^{-1},x^{-1},y^{-1})=s^{2n-2}x^{n-1}Z^{z,w}_n(s,x,y)
\eea
Also (CCW for counterclockwise and CW for clockwise)
\bea
G^{z,w}_n(s,x,y)
={1\over 2\pi i}\oint_{CCW} dz\left(1-{1\over z^2}\right)Z^{z,w}_n(s,x,z^2){1\over z-\sqrt{y}}
\eea
We will study $\sqrt{y}G^{z,w}_n(s,x,y)$ which can be written in two equivalent ways
\bea
\sqrt{y}G^{z,w}_n(s,x,y)
&=&{1\over 2\pi i}\oint_{CCW} dz\left(1-{1\over z^2}\right)Z^{z,w}_n(s,x,z^2){\sqrt{y}\over z-\sqrt{y}}\cr
&=&{1\over 2\pi i}\oint_{CCW} dz\left(1-{1\over z^2}\right)Z^{z,w}_n(s,x,z^2){z\over z-\sqrt{y}}
\eea
Both of the representations will be needed below.
Now, study
\bea
{1\over \sqrt{y}}G^{z,w}_n(s^{-1},x^{-1},y^{-1})
&=&{1\over 2\pi i}\oint_{CCW} dz\left(1-{1\over z^2}\right)Z^{z,w}_n(s^{-1},x^{-1},z^2)
{{1\over\sqrt{y}}\over z-{1\over\sqrt{y}}}\cr
&=&{1\over 2\pi i}\oint_{CCW} dz\left(1-{1\over z^2}\right)Z^{z,w}_n(s^{-1},x^{-1},z^2){1\over \sqrt{y} z-1}
\eea
Now change integration variables from $z$ to $w={1\over z}$ to find
\bea
{1\over \sqrt{y}}G_n^{z,w}(s^{-1},x^{-1},y^{-1})
&=&-{1\over 2\pi i}\oint_{CW} {dw\over w^2} \left(1-w^2\right)Z^{z,w}_n(s^{-1},x^{-1},w^{-2})
{w\over \sqrt{y} -w}\cr
&=&{s^{2n-2}x^{n-1}\over 2\pi i}\oint_{CCW} dw \left(1-{1\over w^2}\right)Z^{z,w}_n(s,x,w)
{w\over w-\sqrt{y}}\cr
&=&s^{2n-2}x^{n-1}\sqrt{y}G^{z,w}_n(s,x,y)
\eea

\subsection{ Gorenstein, Calabi-Yau and top-forms} 

In this section we would like to return to the issue of the Calabi-Yau property for the permutation orbifolds relevant
for the combinatorics of the primaries.
Stanley's  theorem\cite{stanley} tells us that a Cohen Macauly ring that is an integral domain and has a palindromic
Hilbert series, is a Gorenstein ring.
Further, since our rings are defined over an affine space the canonical bundle in this case is trivial, establishing the 
Calabi-Yau property.
According to \cite{HR}, the rings that we consider are Cohen Macaulay because they are the quotient of a Noetherian 
ring $(\mC^2)^n /\mC^2 $ by a reductive group $S_n$. 
However, in general, the relevant rings are not an integral domain. 
It is therefore not clear that we can apply Stanley's theorem to conclude that our permutation orbifolds are Calabi-Yau.

An alternative approach to demonstrating the Calabi-Yau property, is to construct a nowhere vanishing top form.  
To motivate the general formula, it is useful to start with some simple cases. 
For $n=2$ the top form
\bea
\Omega^{ (n-1)} ( dz ) = dz_{12} = dz_1 - dz_2
\eea
is clearly a translation invariant form on $\mC^2$ so it is clearly a top form on the quotient $ \mC^2/ \mC$. 
It is odd under $ S_2$.  
For $n=3$, a translation invariant, $S_n$-odd top form is given by
\bea 
\Omega^{ (n-1)} ( dz ) =  dz_{12} \wedge dz_{ 23}  =  dz_1 \wedge dz_2 -  dz_1 \wedge dz_3 + dz_2 \wedge dz_3   
\eea
For general $n$,
we have 
\bea 
\Omega^{ (n-1)} ( dz ) & = &   dz_{12}\wedge  dz_{ 23}  \wedge \cdots \wedge d z_{ n-1 , n } \cr 
& = & \sum_{ k =1 }^n   I_{ \partial_k } dz_1 \wedge dz_2 \wedge \cdots \wedge dz_n  
\eea
The operator $I_{ \partial_k } $ removes the $dz_k $ in the $n$-form 
and leaves an $ (n-1) $-form, with a sign $ (-1)^{k-1}$. 
In terms of these, the top forms for the orbifolds relevant for the extremal primary problem are 
\bea 
\Omega^{ (n-1)} ( dz ) \wedge \Omega^{ (n-1)} ( dw ) 
\eea

\section{\label{sec:SB}  Vector Model Primaries : Symmetry breaking $S_{2n} \rightarrow S_n[S_2] $ }

Up to now we have considered a single real scalar field.
However, the methods we have developed readily apply in more general settings.
For applications to holography\cite{hs}, it is natural to consider the free gauged $O(N)$ vector model, conjectured to be
dual to higher spin gravity\cite{vas}.
The scalar field is now an $O(N)$ vector and primaries must be $O(N)$ gauge invariants.
In this section we will explain how the techniques we have developed in this article apply to the counting
and construction of primaries in the gauged $O(N)$ vector model.

To obtain a gauge invariant, all vector indices must be contracted.
Thus, to construct a primary, we now distribute the derivatives among
\begin{eqnarray} 
\phi_{ I_1} \phi_{ I_1} \phi_{ I_2} \phi_{ I_2} \cdots \phi_{ I_n } \phi_{ I_n } 
\end{eqnarray}
where the vector indices $I_a$ are summed from $1$ to $N$.
We no longer have an $S_{2n}$ symmetry acting to swap the bosonic fields.
The symmetry is broken to a smaller group which can swap the fields in a given contracted pair, or it can swap the pairs. 
This symmetry group is the wreath product $S_n [S_2]$.
Thus, we don't want to project $V_+^{\otimes 2n}$ onto the trivial of $S_{2n}$ (i.e. $\Sym (V_+^{\otimes 2n})$),
we rather want to project onto the trivial of $S_n [S_2]$. We will restrict attention to the case where $ 2n < N$. This avoids 
subtleties due to finite $N$ relations, associated with the stringy exclusion principle in the context of matrix invariants. 
These can be dealt with using a Young diagram basis, which is left for a future discussion. 

We know the character for the fundamental representation $V_{+}$ of $SO(4,2)$. 
To repeat the analysis we carried out for the free scalar, we need the character for the tensor product of $2n$ fields,
after projecting to the trivial of $S_n[S_2]$.
This gives 
\begin{eqnarray} 
\chi_{{\cal H}_n}(s,x,y)={1\over 2^n n! } \sum_{\sigma\in S_n[S_2]}   {\rm Tr}_{V^{\otimes 2n}} (\sigma  M^{\otimes 2n}) 
\end{eqnarray}
where $M$ is again given by $ s^{\Delta} x^{J_3^L} y^{J_3^R} $. 
This is equal to 
\begin{eqnarray} 
\chi_{{\cal H}_n}(s,x,y)=\sum_{ p \vdash 2n } Z^{ S_n [ S_2] }_{ p }  \prod_{ i =1} ( {\rm Tr} M^i )^{ p_i } 
= \sum_{ p \vdash 2n }  Z^{ S_n [ S_2] }_{ p }\prod_{ i =1}  ( \sum_{a} m_a^i )^{ p_i }
\end{eqnarray}
where $m_a$ are the eigenvalues of $M$ and $Z^{S_n[S_2]}_{p}$ is the cycle index, which gives the number of
permutations in $S_n[S_2]$ with cycle structure specified by $ p_i$. 
The generating function for these cycle indices is known (see e.g. \cite{Cameron}) and can be used to find the following generating function 
for the characters
\begin{align}\label{genfunON} 
 \sum_{n=0}^{ \infty} t^n {\rm Tr}_{{\cal H}_n} (M)  & =  
\prod_{ a } { 1 \over \sqrt{1 - t  m_a^2 }  } \prod_{ a \ne  b } { 1 \over \sqrt{ ( 1 -  t m_a m_b )}  } \cr 
& = \prod_{ a } { 1 \over \sqrt{1 - t  m_a^2}} \prod_{ a > b } { 1 \over ( 1 -  t m_a m_b )  } 
\end{align} 
We can now argue as we did in section \ref{sec:Counting}.
Using the known eigenvalues of $M$, the generalization of (\ref{forchi}) is given by 
\begin{align} 
{\cal Z} ( s , x , y ) &=  \sum_{n=0}^{ \infty} t^n \chi_{{\cal H}_n} (s,x,y)\cr
&=\prod_{ q=0}^{ \infty} \prod_{ a = - { q\over 2} }^{ q \over  2 } \prod_{ b = - { q \over 2 }}^{ q \over 2 } 
{ 1 \over \sqrt{1  - t s^{ 2q+2} x^{ 2a } y^{ 2b }}}  \cr 
& 
\times \prod_{  q_2 = 0 }^{ \infty} \prod_{ a_2 = -  {  q_2 \over 2 }}^{ q_2 \over 2}  \prod_{ b_2 = - { q_2 \over 2 } }^{ q_2 \over 2 } 
 \prod_{ ( q_1 , a_1 , b_1 ) < ( q_2 , a_2 , b_2 ) }
{ 1 \over  ( 1 - t s^{ q_1 + q_2+2 }  x^{  a_1 +a_2 } y^{ b_1 + b_2 } ) }
\end{align}
This can be simplified further.
We can order the triples $ ( q , a , b )$ as follows: The inequality $ ( q_1 , a_1 , b_1 ) <  ( q_2 , a_2 , b_2 )$
means: $ q_1 <  q_2 $ or $ q_1 = q_2$,  $a_1 < a_2 $, or $ q_1 = q_2 , a_1 = a_2 , b_1 < b_2$. 
Alternatively, we can write 
\begin{eqnarray} 
&& {\cal Z} ( s , x , y ) =  \prod_{ q=0}^{ \infty} \prod_{ a = - { q\over 2} }^{ q \over  2 } \prod_{ b = - { q \over 2 }}^{ q \over 2 } { 1 \over \sqrt{ ( 1  - t s^{ 2q+2} x^{ 2a } y^{ 2b } )} } \cr
&&\times\prod_{  q_2 = 0 }^{ \infty} \prod_{ a_2 = -  {  q_2 \over 2 }}^{ q_2 \over 2}  \prod_{ b_2 = - { q_2 \over 2 } }^{ q_2 \over 2 }  \prod_{  q_1 = 0 }^{ \infty} \prod_{ a_1 = -  {  q_1 \over 2 }}^{ q_1 \over 2}  \prod_{ b_1 = - { q_1 \over 2 } }^{ q_1 \over 2 }
{ 1 \over  \sqrt{  ( 1 - t s^{ q_1 + q_2+2 }  x^{  a_1 +a_2 } y^{ b_1 + b_2 } ) } }  \cr 
 && 
\end{eqnarray}
We can now define the generating function (here we take $n>1$ to avoid complications with null states)
\bea
   G_{2n}^{O(N)}(s,x,y)=\sum_{d=0}^\infty\sum_{j_1,j_2}N^{O(N)}_{[2n+d,j_1,j_2]}s^{2n+d} x^{j_1} y^{j_2}
\eea
which is given by
\bea
  G_{2n}^{O(N)}(s,x,y)=\left[ (1-{1\over x})(1-{1\over y})Z_{2n}(s,x,y)\right]_{\ge}
\eea
where
\bea
Z_{2n}(s,x,y)
&=&\chi_{{\cal H}_n} (s,x,y)(1-s\sqrt{xy})(1-s\sqrt{x\over y})(1-s\sqrt{y\over x})(1-{s\over\sqrt{xy}})\cr
&=&\sum_{d=0}^\infty\sum_{j_1,j_2}N^{O(N)}_{[2n+d,j_1,j_2]}s^{2n+d} \chi_{j_1}(x)\chi_{j_2}(y)
\eea
For $n=1$ we need to subtract out the null states that are present since the primaries being counted include
conserved higher spin currents.

We can again specialize to the counting of extremal primaries.
For example, the leading twist primaries are counted by $G_{2n}^{O(N),{\rm max}}(s,x,y)$ where
\bea
\sum_{n=0}^{\infty} t^n G_{2n}^{O(N),{\rm max}}(s,x,y)
=\sum_{n=0}^{\infty} t^n (1-s\sqrt{xy})\chi_{2n}^{O(N),{\rm max}}(s,x,y)
\eea
\begin{eqnarray} 
&&\sum_{n=0}^\infty \chi_{2n}^{O(N),{\rm max}}(s,x,y) t^n =\prod_{q=0}^{\infty}
{1\over\sqrt{1- ts^{ 2q+2} x^{q} y^{q}}}\prod_{q_1,q_2=0}^{\infty}
{1\over\sqrt{1- t s^{ q_1+q_2+2}  x^{q_1 +q_2\over 2} y^{q_1+q_2\over 2}}}  \cr 
 && 
\end{eqnarray}
It is now straightforward to obtain the Hilbert series for leading twist primaries built using 4 fields
\bea
  G^{O(N),{\rm max}}_4(s,x,y)
&=&{s^4 (1-s^6 x^{3}y^{3})\over (1-s^2 xy)^2 (1-s^3 x^{3\over 2}y^{3\over 2})(1-s^4 x^2 y^2)}
\eea
This shows that there are 4 generators and a single relation, that this space of operators is a complete
intersection and it is 3 dimensional.
In a similar way we have
\bea
  G^{O(N),{\rm max}}_6(s,x,y)={s^6 (1-s\sqrt{xy}+s^3 x^{3\over 2}y^{3\over 2}
-s^7 x^{7\over 2}y^{7\over 2}+s^9 x^{9\over 2}y^{9\over 2}-s^{10} x^5 y^5)
\over 
(1-s\sqrt{xy})(1-s^2 xy)^2 (1-s^3 x^{3\over 2}y^{3\over 2})(1-s^4 x^2 y^2)(1-s^6 x^3 y^3)}
\eea
The Hilbert series for these primaries are again palindromic.
For the case of one-complex variable that we are discussing,  we have 
\bea 
G^{O(N),{\rm max}}_{2n} ( q ) =  s^{2n}\sum_{ \substack { \Lambda \vdash 2n \\ \Lambda even  }  } Z_{SH} (q,\Lambda )  
\eea
Using this formula and (\ref{PalZ}) we find
\bea 
G^{O(N),{\rm max}}_{2n} ( q^{-1} ) && =s^{2n}
  \sum_{ \substack { \Lambda \vdash 2n \\ \Lambda even  }  } Z_{SH}  ( q^{-1}   , \Lambda) 
   \cr  
&& = - q^{ 2n-1} s^{ 2n }  \sum_{ \substack { \Lambda \vdash 2n \\ \Lambda even  }  } Z_{SH}  ( q   , \Lambda^T) 
\cr 
&& = - ( q )^{2 n-1}  G^{O(N),{\rm max}} ( q  ) 
\eea
This demonstrates the palindromy property for the Hilbert series associated to the orbifold 
\bea 
( \mC )^{ 2n }   / (  \mC \times S_n [S_2] ) 
\eea

Now consider the two complex variable case.
\bea
{\cal Z} ( s , x , y )&=&\prod_{q=0}^\infty{1\over \sqrt{1- t\,\, s^{ 2q+2} x^{q} y^{q}}}
\prod_{q_1,q_2=0}^{\infty}{1\over \sqrt{1- t\,\, s^{ q_1+q_2+2}  x^{q_1 +q_2\over 2} y^{q_1+q_2\over 2}}} \cr
&=&\sum_{t=0}^\infty t^n\chi^{z,w}_{{\cal H}_n}(s,x,y)
\eea
it is natural to consider the generating functions
\bea
   Z_n^{O(N),zw}(s,x,y)=(1-s\sqrt{xy})(1-s\sqrt{x\over y})\chi^{z,w}_{{\cal H}_n}(s,x,y)
\eea
and
\bea
G_n^{O(N),zw}=\left[ \left(1-{1\over y}\right)Z_n^{O(N),zw}(s,x,y)\right]_{\ge}
\eea
A straightforward computation gives
\bea
  Z^{O(N),zw}_4(s,x,y)=
{g(s,x,y)\over (1-s\sqrt{x\over y})^4(1-s\sqrt{xy})^4 (1+s\sqrt{x\over y})^2 (1+s\sqrt{xy})^2
(1+s^2 {x\over y}) (1+s^2 x y)}\cr\label{VecZ}
\eea
where
\bea
g(s,x,y)&=&s^4 \Big(1-(s\sqrt{x}+s^3x^{3\over 2}+s^5x^{5\over 7}+s^7x^{7\over 2})(\sqrt{y}+{1\over\sqrt{y}})
+(s^8x^4+s^4x^2+2s^2x+2s^6x^3)\cr
&&+(s^4x^2+s^2x+s^6x^3)(y+1+{1\over y})\Big)
\eea
This result can be recovered by using the generating function
\bea
s^4 \sum_{\Lambda_1,\Lambda_2}
(C({\tiny \yng(4)},\Lambda_1,\Lambda_2)+C({\tiny \yng(2,2)},\Lambda_1,\Lambda_2)) 
Z_{SH}(\Lambda_1, s\sqrt{xy})Z_{SH}(\Lambda_2, s\sqrt{x\over y})\label{vecZ}
\eea
Recall that
\bea 
 Z_{SH}(\Lambda,q)=(1-q) q^{\sum_i c_i (c_i -1)/2}\prod_b {1\over (1-q^{ h_b})}     
\eea
Formula (\ref{vecZ}) is a consequence of the fact that an irrep $\Lambda$ of $S_{2n}$ contains the trivial of $S_n[S_2]$ with multiplicity $1$.
For the example given above, using the fact that the non-zero terms are
\bea
C({\tiny \yng(4)},\Lambda_1,\Lambda_2) =\delta_{\Lambda_1,\Lambda_2}\cr
C({\tiny \yng(4)},{\tiny \yng(2,2)},{\tiny \yng(2,2)})=C({\tiny \yng(2,2)},{\tiny \yng(4)},{\tiny \yng(2,2)})=1\cr
C({\tiny \yng(3,1)},{\tiny \yng(3,1)},{\tiny \yng(2,2)})=C({\tiny \yng(2,1,1)},{\tiny \yng(2,1,1)},{\tiny \yng(2,2)})=1\cr
C({\tiny \yng(2,1,1)},{\tiny \yng(3,1)},{\tiny \yng(2,2)})=C({\tiny \yng(3,1)},{\tiny \yng(2,1,1)},{\tiny \yng(2,2)})=1\cr
C({\tiny \yng(1,1,1,1)},{\tiny \yng(2,2)},{\tiny \yng(2,2)})
=C({\tiny \yng(2,2)},{\tiny \yng(1,1,1,1)},{\tiny \yng(2,2)})=1\cr
C({\tiny \yng(2,2)},{\tiny \yng(2,2)},{\tiny \yng(2,2)})=1
\eea
we obtain complete agreement between (\ref{VecZ}) and (\ref{vecZ}).
The geometries associated to $Z_{2n}^{O(N),zw}(s,x,y)$ are 
\bea 
{    ( \mC^{2} )^{ 2n }   \over  ( \mC^2 \times  S_n [ S_2 ]  ) }
\eea
and, after we impose the $G_+$ condition, the geometries for $G_{2n}^{O(N),{\rm zw}}(s,x,y)$ are
\bea 
{   ( \mC^{2} )^{ 2n }  \over  ( G_+ \times S_n [ S_2 ] )  }
\eea
$G_+$ is the unipotent  group of upper triangular  $ 2 \times 2 $ matrices with $1$ on the diagonal. 
For the 2-complex variables case, we have the Hilbert series 
\bea 
Z^{O(N),zw}_n ( q_1 , q_2 ) = s^{2n} \sum_{ \lambda_1 , \Lambda_2  \vdash  2 n } 
  \sum_{ \substack { \Lambda \vdash 2n \\ \Lambda even  }  } C ( \Lambda_1 , \Lambda_2 , \Lambda ) Z_{SH}  ( q_1  , \Lambda_1 ) Z_{ SH} ( q_2 , \Lambda_2 ) 
\eea
where $ C ( R , S , T )$ is the Kronecker coefficient giving the number of $S_n$ invariants in the tensor 
product of three irreps $ R , S  , T $ of $S_n$. 
Applying the inversion 
\bea 
Z^{O(N),zw}_{2n}  ( q_1^{-1}  , q_2^{ -1  }  )  & = & s^{2n} \sum_{ \lambda_1 , \Lambda_2  \vdash  2 n } 
  \sum_{ \substack { \Lambda \vdash 2n \\ \Lambda \rm { even}   }  } C ( \Lambda_1 , \Lambda_2 , \Lambda ) Z_{SH}  ( q_1^{-1}   , \Lambda_1 ) Z_{ SH} ( q_2^{-1 }  , \Lambda_2 ) \cr  
  & = & s^{2n}( q_1 q_2)^{ 2n-1} \sum_{ \lambda_1 , \Lambda_2  \vdash  2 n } 
  \sum_{ \substack { \Lambda \vdash 2n \\ \Lambda \rm { even}   }  } C ( \Lambda_1 , \Lambda_2 , \Lambda ) Z_{SH}  ( q_1   , \Lambda_1^T  ) Z_{ SH} ( q_2  , \Lambda_2^T  ) \cr 
  & = & s^{2n} ( q_1 q_2)^{ 2n-2} \sum_{ \lambda_1 , \Lambda_2  \vdash  2 n } 
  \sum_{ \substack { \Lambda \vdash 2n \\ \Lambda \rm { even}   }  } C ( \Lambda_1^T  , \Lambda_2^T , \Lambda ) Z_{SH}  ( q_1   , \Lambda_1  ) Z_{ SH} ( q_2  , \Lambda_2  ) \cr 
  & = & s^{2n} ( q_1 q_2)^{ 2n-1} \sum_{ \lambda_1 , \Lambda_2  \vdash  2 n } 
  \sum_{ \substack { \Lambda \vdash 2n \\ \Lambda \rm { even}   }  } C ( \Lambda_1  , \Lambda_2 , \Lambda ) Z_{SH}  ( q_1   , \Lambda_1  ) Z_{ SH} ( q_2  , \Lambda_2 ) \cr 
  & = & ( q_1 q_2)^{ 2n-1}  Z^{O(N),zw}_n ( q_1 , q_2 ) \label{vecpal}
\eea
In going from the second to third line, we renamed $ \Lambda_1 \rightarrow \Lambda_1^T,$  
$\Lambda_2 \rightarrow \Lambda_2^T $. 
In going from the third to fourth line, we used an invariance of the Kronecker multiplicity 
\bea 
C ( \Lambda_1 , \Lambda_2 , \Lambda ) = C ( \Lambda_1^T , \Lambda_2^T , \Lambda ) 
\eea
which follows from 
\bea
C ( \Lambda_1 , \Lambda_2 , \Lambda )  =  { 1 \over ( 2n)! } \sum_{ \sigma \in S_{ 2n } } \chi_{ \Lambda_1 } ( \sigma ) \chi_{ \Lambda_2 } ( \sigma ) \chi_{ \Lambda } ( \sigma ) 
\eea
and 
\bea 
\chi_{ \Lambda^T  } ( \sigma ) = ( -1)^{ \sigma } \chi_{ \Lambda } ( \sigma )  
\eea
where $ ( -1)^{ \sigma } $ is the parity of $ \sigma $. 
The formula (\ref{vecpal})  demonstrates that the palindromy property of the 
 Hilbert series for the counting of vector model primaries. 

\section{\label{sec:MM} Matrix Model Primaries}

Another interesting generalization of the single real scalar field, is to a matrix scalar.
We gauge the free theory.
The net effect is that we look for primary operators with all indices contracted.
There are many ways that the indices can be contracted, corresponding to the different possible multitrace
structures that can be written down.
Thus, generalizing to the matrix scalar introduces an interesting non-trivial structure to the problem.

The large $N$ counting of gauge invariant functions of a single matrix, is achieved by integrating\cite{Dolan:2007rq}
\bea 
\cZ ( x ) =   \int dU e^{ \sum_i { x^i \over ^i } ( tr U )^i tr (U^{\dagger})^i } 
=\prod_{i=1}^\infty { 1 \over ( 1 - x^i ) } 
\eea
For multi-matrices, the large $N$ counting is\cite{Dolan:2007rq}
\bea 
   \cZ ( x_i) =  \int dU e^{ \sum_i { ( \sum_a x_a^i ) \over ^i } ( tr U )^i tr (U^{\dagger})^i }
=\prod_{i=1}^\infty { 1 \over ( 1 - \sum_{a=1}^M x_a^i ) } 
\eea
where $M$ is the number of matrices in the model.
Specializing to the 2-matrix case, this is
\bea 
   \cZ ( x , y )=\prod_{i=1}^\infty { 1 \over ( 1 - x^i - y^i ) } 
\eea

For the matrix scalar, we have matrix fields 
\begin{eqnarray} 
   \partial_{l,m} \phi^i_j 
\end{eqnarray} 
$l$ denotes a symmetric traceless irrep of $SO(4)$ and $m$ runs over the states in this irrep.
There are known methods that can be used to write diagonal bases for the local operators of this 
theory\cite{BHR2,deMelloKoch:2011vn}.
For the large $N$ counting of gauge invariants built from derivatives of a single matrix, we have \cite{AMMPV0310}
\bea\label{stpt1} 
\cZ ( t,  s , x , y ) = \int dU e^{\sum_{i=1}^\infty \sum_{q=0}^{\infty} \sum_{a_q,b_q=-{q\over 2}}^{q\over 2}  
{(t s^{(1+q)}x^{a_q}y^{b_q})^ i \over i}tr U^i tr ( U^{\dagger})^i}
\eea
Note that this can also be written as
\bea\label{stpt2}  
 \cZ (t,  s, x , y ) = \int dU e^{\sum_{i=1}^\infty { t^i  \over i } \chi_{V^+} ( s^i , x^i, y^i ) tr U^i tr ( U^{\dagger})^i}
\eea
By repeating steps similar to the ones we did for the integral encountered in case of  multi-matrices, we get 
\bea\label{theanswer} 
\cZ (t, s,x,y)&& =\prod_{i=1}^\infty 
{1\over (1-\sum_{q=0}^\infty\sum_{a_q,b_q=-{q\over 2}}^{q\over 2} t^i s^{i+qi} x^{ia_q} y^{ib_q})}
\eea
To simplify this further, we will derive an identity quoted in \cite{Dolan:2007rq}.
The state space of a single scalar $V_+$ is obtained by acting on the ground state with products of the operators 
$P_{\mu}$. This is a 4D irrep of $SO(4)=SU(2) \times SU(2)$ with spins $(1/2, 1/2)$. The equation of motion says 
that $P_{\mu}P_{\mu}$ acting on the ground state is zero. An immediate consequence is that the  
independent states in $V_+$ generated by $q$ copies of $P$ transform as the symmetric traceless irrep of $SO(4)$,
corresponding to the Young diagram with a single row of length $q$. This irrep of $SO(4)$ is the  $(q/2,q/2)$ 
irrep of $SU(2)\times SU(2)$. It immediately follows that
\bea 
\chi_{ V_+ } ( s , x , y )  && = tr_{ V_+} ( s^{ D } x^{ J_L } y^{ J_R} )  \cr 
 &&  = s \sum_{ q =0}^\infty s^q  \chi_{ q/2} ( x ) \chi_{ q/2} ( y ) \cr 
 && = s \sum_{ q=0}^{ \infty} s^q \sum_{ a_q = - q/2}^{ q/2} x^{a_q}  \sum_{ b_q = - q/2 }^{ q/2}  y^{ b_q} 
\eea
This character was used above in (\ref{stpt1}) and (\ref{stpt2}). 
The state space obtained by acting with all the $P_\mu$'s, without setting $P_\mu P_\mu=0$ has character 
\bea 
\chi_{\tilde V_+} (s,x,y) = tr_{\tilde V_+} s^D x^{J_L} y^{J_R} =  s \sum_{p=0}^\infty 
\sum_{q =0}^\infty s^{2p} s^q \chi_{q/2} (x) \chi_{q/2} ( y ) 
\eea
The $p$ summation is over the number of powers of $P^2$. A basis in $ \tilde V_+$ can be given by multiplying 
powers of $P^2$ with traceless products. Doing  the sum over $p$, we find 
\bea 
\chi_{ \widetilde V_+} ( s , x, y )  = s ( 1- s^2)^{-1} \chi_{ V^+} ( s , x, y ) 
\eea
so that 
\bea 
\chi_{ V^+} ( s, x, y ) = ( 1 - s^2 ) s^{-1}  \chi_{ \widetilde V_+} ( s , x, y )  
\eea
Now by thinking about $\widetilde V_+$ as isomorphic to the Fock space generated by four oscillators 
$P_{\mu}$ (which transform in the  $(1/2 ,1/2)$ of $SU(2)\times SU(2)$) it is evident that
\bea 
\chi_{ \widetilde V_+ } ( s , x, y ) = { s \over ( 1- s \sqrt{ xy} ) ( 1- s \sqrt { x \over y} ) ( 1 - s / \sqrt{ xy} ) ( 1- s \sqrt { y \over x } ) } \equiv s P ( s , x, y ) 
\eea
and so we find
\bea 
\chi_{ V^+} ( s, x, y ) = ( 1- s^2 ) P ( s, x, y )  = s \sum_{ q=0}^{ \infty} s^q \sum_{ a_q = - q/2}^{ q/2} x^{a_q}  \sum_{ b_q = - q/2 }^{ q/2}  y^{ b_q} 
\eea
Thus, we have the identity
\bea 
 \sum_{ q=0}^{ \infty} s^q \sum_{ a_q = - q/2}^{ q/2} x^{a_q}  \sum_{ b_q = - q/2 }^{ q/2}  y^{ b_q}  
= { - ( s  - s^{-1} ) \over ( 1- s \sqrt{ xy} ) ( 1- s \sqrt { x \over y} ) ( 1 - s / \sqrt{ xy} ) ( 1- s \sqrt { y \over x } ) }
\eea
Using this identity, we can now rewrite (\ref{theanswer}) as
\bea
 \cZ (s,x,y)  &=& \prod_{i=1}^\infty \left (1+{(ts)^i(s^i-s^{-i})\over (1-s^i\sqrt{x^i y^i})(1-s^i\sqrt{x^i \over y^i}) 
(1 -  s^i \sqrt { y^i \over x^i } ) ( 1 - s^i /\sqrt{ x^i y^i } ) } \right )^{-1}  \cr 
                 &=& \sum_{n=0}^\infty t^n \chi_n (s,x,y)
\eea
As we did above, we can define two primary generating functions as follows
\bea
Z_n (s,x,y)
&=&\sum_\Delta\sum_{j_1,j_2}\cN^{(n)}_{[\Delta,j_1,j_2]}s^{\Delta} \chi_{j_1}(x)\chi_{j_2}(y)\cr
&=&\chi_n (s,x,y)
(1-s\sqrt{xy})(1-s\sqrt{x\over y})(1-s\sqrt{y\over x})(1-{s\over \sqrt{ xy}})
\eea
and
\bea
G_n (s,x,y)&=&\sum_\Delta\sum_{j_1,j_2}\cN^{(n)}_{[\Delta,j_1,j_2]}s^{\Delta} x^{j_1}y^{j_2}\cr
&=&\left[
\left( 1-{1\over x}\right)\left( 1-{1\over y}\right)Z_n (s,x,y)\right]_{\ge}
\eea
Here $\cN^{(n)}_{[\Delta,j_1,j_2]}$ counts the number of primaries of dimension $\Delta$ and spins $(j_1,j_2)$ that can be constructed using $n$ matrix fields.
We can again specialize the counting to counting leading twist primaries, or to count extremal primaries. 
The relevant generating function for the counting of extremal primaries is given by
\bea 
Z_n^{zw}(s,x,y) =s^n \sum_{\Lambda_1,\Lambda_2 \vdash n}\sum_{R \Lambda \vdash n}
Z_{ SH} ( s\sqrt{xy} , \Lambda_1 )  Z_{ SH} ( s\sqrt{x\over y},\Lambda_2)  C(\Lambda_1,\Lambda_2,\Lambda)  
C(R,R,\Lambda) \cr
\eea
This follows from the general counting of matrix gauge invariants in the case where the matrices $X_a$ transform 
under some global symmetry group $G$, given in \cite{BHR2}. 
The resulting Hilbert series, for $n=3$, is

\vfill\eject

\bea
Z_3^{zw} =
\frac{s^3 \Upsilon (s,x,y)}
{(1-s\sqrt{x\over y})^2 (1+s \sqrt{x\over y}) (-1+s \sqrt{xy})^2 (1+s \sqrt{xy}) \left(s^2 {x\over y}
+s \sqrt{x\over y}+1\right) 
\left(1+s \sqrt{x y}+s^2 x y\right)}\cr
\eea
\bea
\Upsilon (s,x,y)=3+3 s^6 x^3 +(s \sqrt{x}  +s^5 x^{5\over 2} ) ({1\over \sqrt{y}}+\sqrt{y})+
(s^2 x +s^4 x^2) ({1\over y}+5+y)\cr
+s^3 x^{3\over 2} ({1\over y^{3\over 2}}+ {5\over\sqrt{y}}+5\sqrt{y}+y^{3\over 2})
\eea
This counts the total number of primaries we can build from 3 matrix fields.
We can refine this counting by specifying the trace structure.
Schematically, the primaries we study have the form
\bea
   {\cal O}=\sum_{\vec n,\vec m}c_{\vec n\,\,\vec m}
\partial_{z_1}^{n_1}\partial_{w_1}^{m_1}\phi^{i_1}_{i_{\sigma(1)}}
\partial_{z_2}^{n_2}\partial_{w_2}^{m_2}\phi^{i_2}_{i_{\sigma(2)}}
\partial_{z_3}^{n_3}\partial_{w_3}^{m_3}\phi^{i_3}_{i_{\sigma(3)}}\Big|_{z_k=z,w_k=w}
\eea
i.e. they are specified by allowing derivatives to act on some gauge invariant operator specified by the permutation $\sigma\in S_n$.
After we translate to the polynomial language, primaries are specified by polynomials in $n$ variables $z_i$ and $w_i$,
as well as by the trace structure, i.e. they are functions on the space
\bea
{   (  \mC^2 )^n \over \mC^2 }  \times S_n 
\eea
These functions have to be invariant under an action of  $ \gamma \in S_n$ 
\bea
 \gamma :   (w_{ I } ,z_{ J } ,\sigma) \rightarrow  ( w_{ \gamma(I)} ,z_{ \gamma (I)} ,\gamma^{-1}\sigma\gamma)\qquad \gamma\in S_n
\eea
Modding out by this symmetry we find the primaries are functions on the space
\bea
   (  (\mC^2 )^{ n }  \times S_n )  \over (  \mC^2 \times S_n ) 
\eea
We can also obtain a description by fixing a specific permutation, and then dividing by those permutations $\gamma$
that fix $\sigma$.
Lets work out this description for $n=3$.
For primaries obtained by acting with derivatives on ${\rm Tr}(\phi )^3$, $\sigma = (1)(2)(3)$
which is left invariant by $\gamma\in S_3$. Thus, we need to consider
\bea
{ (\mC^2 )^3  \over (  \mC^2 \times S_3 )  } 
\eea
We need to project to the trivial of $S_3$ and hence
\bea
Z^{zw}_{ ( \Tr \phi)^3}
&=&s^3\sum_{\Lambda\vdash 3}Z_{ SH} ( s\sqrt{xy} , \Lambda) Z_{ SH} ( s\sqrt{x\over y},\Lambda)\cr
&=&\frac{s^3\left(1+s^2 x +s^4 x^2+s^6 x^3 +s^3 x^{3\over 2} \left({1\over\sqrt{y}}+\sqrt{y}\right)\right)}
{(1-s\sqrt{xy})^2 (1+s \sqrt{xy}) (-1+s\sqrt{xy})^2 (1+s\sqrt{xy}) 
\left(s^2 {x\over y}+s \sqrt{x\over y}+1\right) \left(1+s\sqrt{xy}+s^2 x y\right)}\cr
&&
\eea
For primaries obtained by acting with derivatives on $( {\rm Tr}\phi^2){\rm Tr}(\phi )$, we can choose
$\sigma =(12)(3)$ which is left invariant by $S_2\times S_1$. 
Thus, we need to consider
\bea
{ (\mC^2 )^3  \over (  \mC^2 \times S_2 \times S_1  )  } 
\eea
where $S_2$ contains permutations of $(z_1,w_1)$ and $(z_2,w_2)$.
Thus, we need to project to the trivial $({\tiny \yng(2)},{\tiny \yng(1)})$ of the $S_2\times S_1$ subgroup. 
This representation is subduced once by ${\tiny \yng(3)}$ and once by ${\tiny \yng(2,1)}$.
Thus
\bea
Z^{zw}_{ ( \Tr  \phi^2)\Tr(\phi)}&=&
s^3Z_{ SH} ( s\sqrt{xy} , {\tiny \yng(3)})  Z_{ SH} ( s\sqrt{x\over y},{\tiny\yng(3)})+
2s^3Z_{ SH} ( s\sqrt{xy} , {\tiny \yng(2,1)})  Z_{ SH} ( s\sqrt{x\over y},{\tiny\yng(2,1)})\cr
&+&s^3Z_{ SH} ( s\sqrt{xy} , {\tiny \yng(1,1,1)})  Z_{ SH} ( s\sqrt{x\over y},{\tiny\yng(1,1,1)})
+s^3Z_{ SH} ( s\sqrt{xy} , {\tiny \yng(3)})  Z_{ SH} ( s\sqrt{x\over y},{\tiny\yng(2,1)})\cr
&+&s^3Z_{ SH} ( s\sqrt{xy} , {\tiny \yng(2,1)})  Z_{ SH} ( s\sqrt{x\over y},{\tiny\yng(3)})
+s^3Z_{ SH} ( s\sqrt{xy} , {\tiny \yng(2,1)})  Z_{ SH} ( s\sqrt{x\over y},{\tiny\yng(1,1,1)})\cr
&+&s^3Z_{ SH} ( s\sqrt{xy} , {\tiny \yng(1,1,1)}) Z_{ SH} ( s\sqrt{x\over y},{\tiny\yng(2,1)})\cr
&=&\frac{s^3(1+s^2 x)}{(1-s\sqrt{xy})^2 (1+s\sqrt{xy}) (-1+s\sqrt{xy})^2 (1+s\sqrt{xy})}
\eea
For primaries obtained by acting with derivatives on ${\rm Tr}(\phi^3)$, we can take $\sigma=(123)$ which is
left invariant by $Z_3$. Thus, need to consider
\bea
 { (\mC^2 )^3  \over (  \mC^2 \times Z_3 )  } 
\eea
where $Z_3$ is the group comprising $\{1,(123),(132)\}$.
We need to project to the trivial of $Z_3$. 
The trivial of $Z_3$ is subduced once by ${\tiny \yng(3)}$ and once by ${\tiny\yng(1,1,1)}$.
Thus
\bea
Z^{zw}_{( \Tr \phi^3)}&=&
s^3Z_{ SH} ( s\sqrt{xy} , {\tiny \yng(3)})  Z_{ SH} ( s\sqrt{x\over y},{\tiny\yng(3)})+
2s^3Z_{ SH} ( s\sqrt{xy} , {\tiny \yng(2,1)})  Z_{ SH} ( s\sqrt{x\over y},{\tiny\yng(2,1)})\cr
&+&s^3Z_{ SH} ( s\sqrt{xy} , {\tiny \yng(1,1,1)})  Z_{ SH} ( s\sqrt{x\over y},{\tiny\yng(1,1,1)})
+s^3Z_{ SH} ( s\sqrt{xy} , {\tiny \yng(3)})  Z_{ SH} ( s\sqrt{x\over y},{\tiny\yng(1,1,1)})\cr
&+&s^3Z_{ SH} ( s\sqrt{xy} , {\tiny \yng(1,1,1)}) Z_{ SH} ( s\sqrt{x\over y},{\tiny\yng(3)})\cr
&=&{s^3\left(1+s^4 x^2-(s\sqrt{x}+s^3 x^{3\over 2}) ({1\over\sqrt{y}}+\sqrt{y})+ 
   s^2 x({1\over y} + 3 + y)\right)\over (1-s\sqrt{x\over y})^2 (1-s\sqrt{xy})^2 (s^2 {x\over y}+s\sqrt{x\over y}+1) 
(1+s\sqrt{xy}+s^2 xy)}
\eea
Note that
\bea 
Z^{zw}_3=Z^{zw}_{\Tr (\phi)^3}+Z^{zw}_{\Tr (\phi^2)\Tr(\phi)}+Z^{zw}_{\Tr (\phi^3)}
\eea
as it must be.
The permutation quotient geometry which includes all trace structures is 
\bea
{  ( \mC^2 )^n \times S_n  \over  ( \mC^2 \times  S_n )  }  
\eea
This has an $SU(2)$ action. We can again look at functions which are annihilated by $J_+$. 
Let $ G_+$ be the subalgebra of $ GL(2,\mC)$ generated by $J_+$.
The Hilbert series in this case is $G^{zw}_n$. 
The algebra of functions annihilated by $J_+$ corresponds to functions on 
\bea 
{  ( \mC^2 )^n \times S_n  \over  ( \mC^2 \times  S_n \times G_+  )  }  
\eea
It is again possible to establish the palindromic property for the Hilbert series relevant for the matrix case.
In the matrix case, we have the counting function 
\bea 
Z^{zw}_n ( q_1 , q_2 ) = s^n \sum_{ \Lambda_1 , \Lambda_2 \vdash n } \sum_{ R \vdash n } 
C ( \Lambda_1 , \Lambda_2 , \Lambda ) C ( R , R , \Lambda ) Z_{ SH} ( q_1 , \Lambda_1 ) Z_{ SH} ( q_2 , \Lambda_2 ) 
\eea
The symmetry under $ q_1 \leftrightarrow q_2$, equivalently $ x\rightarrow x , y \rightarrow y^{-1} $ is clear. 
Now apply inversion 
\bea 
Z^{zw}_n ( q_1^{-1 }  , q_2^{-1}  )  & = & s^n  \sum_{ \Lambda_1 , \Lambda_2 \vdash n } \sum_{ R \vdash n } 
C ( \Lambda_1 , \Lambda_2 , \Lambda ) C ( R , R , \Lambda ) Z_{ SH} ( q_1^{-1}  , \Lambda_1 ) Z_{ SH} ( q_2^{-1}  , \Lambda_2 ) \cr  
 & = &s^n ( q_1 q_2)^{ n-1} \sum_{ \Lambda_1 , \Lambda_2 \vdash n } \sum_{ R \vdash n } 
C ( \Lambda_1 , \Lambda_2 , \Lambda ) C ( R , R , \Lambda ) Z_{ SH} ( q_1  , \Lambda_1^T  ) Z_{ SH} ( q_2  , \Lambda_2^T  ) \cr 
& = & s^n ( q_1 q_2)^{ n-1} \sum_{ \Lambda_1 , \Lambda_2 \vdash n } \sum_{ R \vdash n } 
C ( \Lambda_1^T  , \Lambda_2^T  , \Lambda ) C ( R , R , \Lambda ) Z_{ SH} ( q_1  , \Lambda_1  ) Z_{ SH} ( q_2  , \Lambda_2 ) \cr 
& = &  s^n ( q_1 q_2)^{ n-1} \sum_{ \Lambda_1 , \Lambda_2 \vdash n } \sum_{ R \vdash n } 
C ( \Lambda_1  , \Lambda_2  , \Lambda ) C ( R , R , \Lambda ) Z_{ SH} ( q_1  , \Lambda_1  ) Z_{ SH} ( q_2  , \Lambda_2 ) \cr 
& = & ( q_1 q_2)^{ n-1} Z^{zw}_n ( q_1 , q_2 ) 
\eea

\section{ Summary and Outlook } 

We  mapped the algebraic problem of constructing primary fields in  the quantum field theory of 
a free scalar field $ \phi$ in four dimensions to one of 
finding polynomial functions on $ ( \mR^4 )^n $ subject to constraints involving Laplace's 
equation on each factor, a condition of invariance under translations by the diagonal $ \mR^4$ and 
an $S_n$ symmetry related to the bosonic statistics of the elementary field (\ref{constraints}).  
By considering holomophic solutions to the Laplacian conditions, we mapped the primary fields to 
functions on the complex orbifold 
\bea
( \mC^2)^n /(  \mC^2 \times S_n ) 
\eea
We showed that this space has a palindromic Hilbert series and is Calabi-Yau. 
We generalized the discussion to the quantum field theory of free vector fields $ \phi_I ( x ) $  in the large $N$ limit 
and found that the orbifold 
\bea 
( \mC^2)^{ 2n }  /(  \mC^2 \times S_n [ S_2]  ) 
\eea
plays an analogous role. We established the palindromy property. 
We then considered the free matrix scalar in four dimensions $ \phi^j_{i} ( x ) $ again in the large $N$ limit. 
The orbifold is now 
\bea 
 ( ( \mC^2)^{n  } \times S_n )   /(  \mC^2 \times S_n  ) 
\eea
We established the palindromy of the Hilbert series

In this paper we have focused on the explicit construction of extremal primary fields. However, the formulation of
the problem of constructing general primary fields given in (\ref{constraints}), as a system of 
equations for harmonic polynomal functions on $ ( \mR^4)^n $, should be useful beyond the extremal sector. 
In this more general case, we have to include non-holomorphic solutions to the harmonic  constraints - solving this simultaneously 
with the symmetry and translation constraints proves surprisingly tricky.  In this case, we do not expect the ring structure of the extremal primaries to survive.  Our preliminary investigations indicate that this most general problem has a graph-theoretic formulation,  which will be interesting to exploit.  
At the level of counting  these primaries, we still have the full expressions for the $so(4,2)$ characters of $Sym^n ( V_+)$ which, once expanded  in terms of irreducible representations, will in principle yield the counting for the general case. However finding 
explicit expressions analogous to (\ref{nis3zw}) or (\ref{G4form}) looks challenging. It would very interesting to explore the possible  application  of the  higher spin symmetries and twistor space variables of \cite{VasilievMulti,GelVas} in shedding light on this problem. It is interesting to note that symmetric group representation theoretic questions close to (but not identical) to  the ones we have used have played a role in the discussion of higher spin symmetries in \cite{GopGab1406}. Some recent mathematical results 
on these symmetric group multiplicities are in \cite{BHH1605}.

A number of immediate generalizations of the
current work  are: free fermions, gauge fields, the free limit of QCD and 
supersymmetric theories. Some of the early constructions of primary fields - in the $SL(2)$ sector which is a special case of 
the extremal operators we considered -  were done in the context of  deep inelastic scattering in QCD (see for example the review \cite{BKM03}). 
It will be fascinating to explore QCD applications  of the holomorphic primaries considered here. The explicit enumeration and 
construction of superconformal primary fields in $N=4$ SYM will give a better understanding  
of the dual $ AdS_5 \times S^5$ background. While the map between branes and geometries in the half-BPS sector of the bulk and  the half-BPS states in $N=4$ SYM\cite{CJR,Dtoy,Lin:2004nb} is reasonably well understood, there are important open problems, most  notably in the sector of 
sixteenth BPS states \cite{GGSM08} but also in the quarter and eighth-BPS sectors (some progress on branes states in these sectors is 
in \cite{CountCons,FuzMod,Mikh2000,BGLM07,Koch:2011hb,deMelloKoch:2012ck,Berenstein:2013md,Berenstein:2014pma}).    A better understanding of operators with derivatives is a step in the direction of a more complete picture of  the duality map in general. The construction of holomorphic primaries for the 1-matrix case should admit, without much diffculty, generalization to 
multi-matrix systems and more generally to quiver theories by combining the methods of the present paper with those of 
\cite{SY,Brown:2007xh,Bhattacharyya:2008rb,Bhattacharyya:2008xy,BHR2,quivcalc,DKN1403}. 
Another natural direction is to consider  correlators involving the extremal primary fields and the determination of 
anomalous dimensions for these fields at the Wilson-Fischer fixed point using the techniques of \cite{Rychkov:2015naa}.

{\vskip 0.5cm}
\noindent
\begin{centerline} 
{\bf Acknowledgements}
\end{centerline} 

This work of RdMK, PR and RR is supported by the South African Research Chairs
Initiative of the Department of Science and Technology and National Research Foundation
as well as funds received from the National Institute for Theoretical Physics (NITheP).
SR is supported by the STFC consolidated grant ST/L000415/1 “String Theory, Gauge Theory \& Duality”
and  a Visiting Professorship at the University of the Witwatersrand, funded by a Simons Foundation 
grant held at the Mandelstam Institute for Theoretical Physics. SR thanks the Galileo Galilei Institute for Theoretical Physics 
for  hospitality and the INFN for partial support during the completion of this work. We are grateful for useful discussions to Alberto  Cazzaniga, Danilo Diaz,  Yang Hui He, Dario Martelli, Vishnu Jejjala, Bogdan Stefanski, Alessandro Torielli.  

\appendix

\section{ Decomposing $\Sym^n (V_+)$ for small values of $n$}

In this Appendix we will discuss an alternative approach to the problem of decomposing $\Sym (V_+^{\otimes n})$ into irreps for some low values of $n$. This approach was developed in detail in  \cite{tftpap}.
The results obtained using the methods outlined in this Appendix are in complete agreement with the results
derived in section \ref{sec:Counting}.
The method of \cite{tftpap} starts with the observation that projection onto the completely symmetrized representation is easily accomplished with the help of Young projectors.
For example, for $n=3$ we have
\bea
\chi_{\Sym^3 (V)}={1\over 6}\left((\chi_{V_+}(s,x,y))^3 +3\chi_{V_+}(s,x,y)\chi_{V_+}(s^2,x^2,y^2)
                                          +2\chi_{V_+}(s^3,x^3,y^3)\right)
\label{sym3}
\eea
Evaluating the right hand side is most easily achieved by using the formula
\bea
  \chi_{V^+}(s^n,x^n,y^n)&&=P(s^n,x^n,y^n)s^n (1-s^{2n})\nn\cr
&&=s^{n}\sum_{q=0}^\infty s^{nq}
\left[\sum_{l=0,1,...}^{\lfloor q/2\rfloor}\chi_{{{qn\over 2}-nl}}(x)-\sum_{l=0,1,...}^{\lfloor (q-1)/2\rfloor}
            \chi_{{{qn\over 2}-nl-1}}(x)\right]\nn\\
&&\quad\times\left[\sum_{l=0,1,...}^{\lfloor q/2\rfloor}\chi_{{{qn\over 2}-nl}}(y)-\sum_{l=0,1,...}^{\lfloor (q-1)/2\rfloor}
            \chi_{{{qn\over 2}-nl-1}}(y)\right]
\label{withoutP}
\eea
We also need an identity which rewrites $\chi_{V^+}(s^n,x^n,y^n)$ as $SU(2)$ characters multiplied by $P(s,x,y)$; these can very easily be translated into ${\cal A}_{[\cdot,\cdot,\cdot]}$s, in the notation of \cite{Dolan:2007rq}. 
This is easily achieved by using the well known product rule for $SU(2)$ characters as well as the identity
\bea
  1&&=P(s,x,y)(1-sx^{1/2}y^{1/2})(1-sx^{1/2}y^{-1/2})(1-sx^{-1/2}y^{1/2})(1-sx^{-1/2}y^{-1/2})\nn\\
   &&=P(s,x,y)\big[1+s^4-s(1+s^2)\chi_{1\over 2}(x)\chi_{1\over 2}(y)+s^2(\chi_1(x)+\chi_1(y))\big]
\eea
A straight forward computation now gives the desired decomposition.
A few examples of the method are given below.
\bea
\chi_{\Sym^3 (V)}
&&={\cal A}_{[3,0,0]}+{\cal A}_{[5,1,1]}+{\cal A}_{[6,{3\over 2},{3\over 2}]}+{\cal A}_{[7,2,2]}+{\cal A}_{[7,0,2]}
+{\cal A}_{[7,2,0]}\nn\\
&&+{\cal A}_{[8,{5\over 2},{5\over 2}]}+{\cal A}_{[8,{3\over 2},{5\over 2}]}+{\cal A}_{[8,{5\over 2},{3\over 2}]} + 2{\cal A}_{[9,3,3]} + {\cal A}_{[9,1,3]}  +{\cal A}_{[9,3,1]}\nn\\
&&+{\cal A}_{[10,{7\over 2},{7\over 2}]}+{\cal A}_{[10,{7\over 2},{5\over 2}]}+{\cal A}_{[10,{5\over 2},{7\over 2}]}
+{\cal A}_{[10,{7\over 2},{3\over 2}]}+{\cal A}_{[10,{3\over 2},{7\over 2}]}+...
\eea
\bea
\chi_{\Sym^4 (V)}&&=
{\cal A}_{[4,0,0]}+{\cal A}_{[6,1,1]}+{\cal A}_{[7,{3\over 2},{3\over 2}]}
+{\cal A}_{[8,0,0]}+{\cal A}_{[8,0,2]}+{\cal A}_{[8,2,0]}+{\cal A}_{[8,1,1]}+2{\cal A}_{[8,2,2]}\nn\\
&&+{\cal A}_{[9,{3 \over 2},{1 \over 2}]}+{\cal A}_{[9,{5 \over 2},{1 \over 2}]}+{\cal A}_{[9,{1 \over 2},{3 \over 2}]}+{\cal A}_{[9,{5 \over 2},{3 \over 2}]}+{\cal A}_{[9,{1 \over 2},{5 \over 2}]}+{\cal A}_{[9,{3 \over 2},{5 \over 2}]}+{\cal A}_{[9,{5 \over 2},{5 \over 2}]}\nn\\
&&+{\cal A}_{[10,0,0]} + 2{\cal A}_{[10,1,1]}+{\cal A}_{[10,2,1]}+2{\cal A}_{[10,3,1]}+{\cal A}_{[10,1,2]}+2{\cal A}_{[10,2,2]}+{\cal A}_{[10,3,2]}\nn\\
&&+2{\cal A}_{[10,1,3]}+{\cal A}_{[10,2,3]}+3{\cal A}_{[10,3,3]}+...
\eea
\bea
\chi_{\Sym^5 (V)}&&=
{\cal A}_{[5,0,0]}+{\cal A}_{[7,1,1]}+{\cal A}_{[8,{3\over 2},{3\over 2}]}+
{\cal A}_{[9,0,0]}+{\cal A}_{[9,1,1]}+{\cal A}_{[9,2,0]}+{\cal A}_{[9,0,2]}+2{\cal A}_{[9,2,2]}\cr
&&+{\cal A}_{[10,{1\over 2},{1\over 2}]}+{\cal A}_{[10,{3\over 2},{1\over 2}]}
+{\cal A}_{[10,{1\over 2},{3\over 2}]}+{\cal A}_{[10,{3\over 2},{3\over 2}]}
+{\cal A}_{[10,{1\over 2},{5\over 2}]}+
{\cal A}_{[10,{5\over 2},{1\over 2}]}+{\cal A}_{[10,{5\over 2},{3\over 2}]}\cr
&&+{\cal A}_{[10,{3\over 2},{5\over 2}]}+2{\cal A}_{[10,{5\over 2},{5\over 2}]}+...
\eea

\section{Generating function of characters}

To count the primaries in the $O(N)$ vector model, we needed explicit expressions for the characters of 
$V_+^{\otimes n}$ projected to the trivial of $S_n[S_2]$. In this Appendix we will derive the generating function
\bea
\cZ (t,Q)=\sum_{n=0}^\infty t^n\chi_{\cH_n}(Q)
\eea 

The generating function of characters for $ {\cal H}_n$, the $S_n[S_2]$ invariant subspace of $ V^{ \otimes 2n }$, is 
\begin{eqnarray}\label{genChar}  
 {\cal Z} ( t , Q  ) &= 
\sum_{ n =0}^{ \infty} { t^n \over 2^n n! } \sum_{ \sigma \in S_n [ S_2] } {\rm tr}_{ V^{ \otimes 2n }  } ( \sigma Q^{ \otimes 2n } )  \cr 
& \sum_{ n=0}^{ \infty} \sum_{ p \vdash 2n } {\cal Z}^{ S_n[S_2]}_{ \vec p } \prod_{ i} ( {\rm tr} Q^i )^{p_i} \cr 
& = \sum_{ n=0}^{ \infty} t^n \sum_{ p \vdash 2n }  {\cal Z}^{ S_n[S_2]}_{ \vec p } \prod_{ i} (\sum_{ a } q_a^i )^{p_i} 
\end{eqnarray} 
Here $ {\cal Z}^{ S_n[S_2]}_{ \vec p } $ is the number of permutations in $ S_n[S_2]$ with cycle structure $ \vec p $, 
divided by the order of $ S_n [ S_2]$. The cycle polynomials are
\begin{eqnarray} 
{\cal Z}^{ S_n[S_2]} ( \vec x ) & = &  \sum_{ p \vdash 2n } {\cal Z}_{ \vec p }^{ S_n[S_2]} \prod_i x_i^{ p_i} 
\end{eqnarray}
The generating function of the cycle polynomials is given by 
\begin{eqnarray}\label{genCyc}  
{\cal Z} ( t , \vec x ) & = &  \sum_{ n =0}^{ \infty} t^n {\cal Z}^{ S_n[S_2]} ( \vec x )  \cr 
    & = & e^{ \sum_{ i=1}^{ \infty} { t^i \over 2i } \left ( x_{ 2i} + x_i^{2}  \right ) }
\end{eqnarray}
Comparing    (\ref{genChar}) and (\ref{genCyc}) we see that 
\begin{align} 
{\cal Z} ( t , Q  ) & = {\cal Z} ( t ,  x_i \rightarrow \sum_a q_a^i ) \cr 
 & = e^{ \sum_{ i=1}^{ \infty} {  t^i \over 2 i } \left (  \sum_a q_a^i )^2 + \sum_a q_a^{ 2i }      \right )}  \cr 
   & = e^{ \sum_{ i=1}^{ \infty} {  t^i \over 2 i } \left (   \sum_a \sum_b q_a^i q_b^i  + \sum_a q_a^{ 2i }      \right ) } \cr 
   & = e^{ \sum_{ a , b } \sum_{ i=1}^{ \infty} { t^i q_a^i q_b^i  \over 2i } +\sum_a  \sum_{ i }{  t^i \over 2i }q_a^{ 2i}} \cr 
   & = e^{  - { 1 \over 2 } \sum_{ a , b } \log ( 1 - t q_a q_b ) - { 1 \over 2 } \sum_{ a } \log ( 1 - t q_a^2 ) } \cr 
   & = \prod_{ a } { 1 \over  \sqrt{ 1- q_a^2}  } \prod_{ a , b } { 1 \over \sqrt { ( 1 - t q_a q_b )}  } \cr 
   & = \prod_{ a } { 1 \over \sqrt{1- q_a^2}} \prod_{ a <  b } { 1 \over  ( 1 - t q_a q_b )  } \cr 
\end{align} 

\section{The Hilbert Series for $Z_3 (s,x,y)$}\label{HS}

In this Appendix we consider the Hilbert series $Z_3 (s,x,y)$ for the counting of extremal primaries built using
3 scalar fields. 
This Hilbert series has a non-trivial numerator
\bea
Z_3^{z,w}=
{s^3 \big(1-s^5 x^{5\over 2}(\sqrt{y}+{1\over \sqrt{y}})
-s^6 x^3 ({1\over y}+1+y)-s^{14}x^7+
s^8 x^4(y+1+{1\over y})+s^9x^{9\over 2}(\sqrt{y}+{1\over \sqrt{y}})\big)
\over
(1-s^2 x y)(1-s^2 x)(1-s^2 {x\over y})(1-s^3 x^{3\over 2} y^{3\over 2})(1-s^3 x^{3\over 2} \sqrt{y})
(1-s^3 {x^{3\over 2}\over \sqrt{y}})(1-{s^3 x^{3\over 2}\over y^{3\over 2}})}
\nonumber
\eea
Our goal in this Appendix is to explain how the numerator of $Z_3 (s,x,y)$ encodes relations between the
generators of the ring as well as relations between those relations.
 
From the denominator of the Hilbert series, we have 7 generators.
We can easily identify them as follows
\begin{eqnarray}
G_1=(z_{12})^{2}+(z_{13})^{2}+(z_{23})^{2} &\leftrightarrow& s^2 xy\cr\cr
G_2=z_{12}w_{12}+z_{13}w_{13}+z_{23}w_{23} &\leftrightarrow& s^2 x\cr\cr
G_3=(w_{12})^{2}+(w_{13})^{2}+(w_{23})^{2} &\leftrightarrow& s^2 {x\over y}\cr\cr
G_4=(z_{13}+z_{23}) (z_{31}+z_{21}) (z_{12}+z_{32}) &\leftrightarrow& s^3 x^{3\over 2}y^{3\over 2}\cr\cr
G_5=(w_{13}+w_{23}) (z_{31}+z_{21}) (z_{12}+z_{32})&\cr
+(z_{13}+z_{23}) (w_{31}+w_{21}) (z_{12}+z_{32})&\cr
+(z_{13}+z_{23}) (z_{31}+z_{21}) (w_{12}+w_{32})
 &\leftrightarrow& s^3 x^{3\over 2}y^{1\over 2}\cr\cr
G_6=(w_{13}+w_{23}) (w_{31}+w_{21}) (z_{12}+z_{32})&\cr
+(z_{13}+z_{23}) (w_{31}+w_{21}) (w_{12}+w_{32})&\cr
+(w_{13}+w_{23}) (z_{31}+z_{21}) (w_{12}+w_{32})
 &\leftrightarrow& s^3 x^{3\over 2}y^{-{1\over 2}}\cr\cr
G_7=(w_{13}+w_{23}) (w_{31}+w_{21}) (w_{12}+w_{32}) 
&\leftrightarrow& s^3 x^{3\over 2}y^{-{3\over 2}}
\end{eqnarray}
From the numerator of the Hilbert series, the terms with a negative sign should correspond to relations between
the generators of the degree given by the monomial. 
From $-s^5 x^{5\over 2}(\sqrt{y}+{1\over \sqrt{y}})-s^6 x^3 ({1\over y}+1+y)-s^{14}x^7$
we have 6 relations.
They are
\bea
\chi_1=3 G_3 G_4-2 G_2 G_5+G_1 G_6=0 &\leftrightarrow& s^5 x^{5\over 2}\sqrt{y}\cr\cr
\chi_2=G_3 G_5-2 G_2 G_6+3 G_1 G_7=0&\leftrightarrow& {s^5 x^{5\over 2}\over \sqrt{y}}\cr\cr
\chi_3=4 G_1 G_2^2-G_1^2 G_3=0&\leftrightarrow& s^6 x^3 y\cr\cr
\chi_4=4 G_2^3-G_1 G_2 G_3 =0&\leftrightarrow& s^6 x^3\cr\cr
\chi_5=4 G_2^2 G_3-G_1 G_3^2=0&\leftrightarrow& {s^6 x^3\over y}\cr\cr
\chi_6=G_2^7-G_1 G_2^5 G_3+{1\over 9} G_2^4 G_5 G_6-G_2^4 G_4 G_7=0&\leftrightarrow& s^{14} x^7 
\eea
Again from the numerator of the Hilbert series, the terms with a positive sign should corresponds to relations between
the relations, again of the degree given by the monomial.
From $s^8 x^4(y+1+{1\over y})+s^9x^{9\over 2}(\sqrt{y}+{1\over \sqrt{y}})$ we have 5 relations among
the relations.
They are
\bea
4\chi_5 G_2+\chi_4 G_3 =0 &\leftrightarrow& {s^8 x^{4}\over y}\cr\cr
\chi_5 G_1 - \chi_3 G_3=0 &\leftrightarrow& s^8 x^{4}\cr\cr
\chi_4 G_1+4 \chi_3 G_2=0 &\leftrightarrow& s^8 x^{4}y\cr\cr
\chi_2 G_2^2-{1\over 4} \chi_2 G_1 G_3- \chi_5 G_5-{1\over 2} \chi_4 G_6-3 \chi_3 G_7=0 
&\leftrightarrow& {s^9 x^{9\over 2}\over\sqrt{y}}\cr\cr
-4\chi_1 G_2^2+\chi_1 G_1 G_3+12 \chi_5 G_4+2 \chi_4 G_5+4 \chi_3 G_6
=0 &\leftrightarrow& s^9 x^{9\over 2}\sqrt{y}
\eea

{} 

\end{document}